\documentclass[prb,reprint,showkeys,superscriptaddress]{revtex4-1}  
\usepackage{graphicx}  
\usepackage{dcolumn}   
\usepackage{bm}        
\usepackage{amssymb}   
\graphicspath{{Pictures/}} 
\usepackage{booktabs}
\usepackage{amsmath,amssymb,color,float,setspace}
\usepackage[utf8]{inputenc}
\usepackage[T1]{fontenc}
\usepackage{titlesec}
\usepackage{booktabs}
\usepackage{booktabs}
\usepackage[bottom]{footmisc}
\usepackage{multirow}
\usepackage{ulem}

\usepackage{hhline}
\newcommand{\red}{\color{red}}

\begin{document}

\title{Altering the magnetic ordering of Fe$_{3}$Ga$_{4}$ via thermal annealing and hydrostatic pressure}
\author{Brandon Wilfong}
\affiliation{Physics Department, United States Naval Academy, Annapolis, MD 20899, USA}
\author{Vaibhav Sharma}
\affiliation{Mechanical and Nuclear Engineering, Virginia Commonwealth University, Richmond, VA 23220, USA}
\author{Jared Naphy}
\affiliation{Physics Department, United States Naval Academy, Annapolis, MD 20899, USA}
\author{Omar Bishop}
\affiliation{Mechanical and Nuclear Engineering, Virginia Commonwealth University, Richmond, VA 23220, USA}
\author{Steven P. Bennett}
\affiliation{Material Science and Technology Division, U.S. Naval Research Laboratory, Washington, DC, 20375, USA}
\author{Joseph Prestigiacomo}
\affiliation{Material Science and Technology Division, U.S. Naval Research Laboratory, Washington, DC, 20375, USA}
\author{Radhika Barua}
\affiliation{Mechanical and Nuclear Engineering, Virginia Commonwealth University, Richmond, VA 23220, USA}
\author{Michelle E. Jamer}
\affiliation{Physics Department, United States Naval Academy, Annapolis, MD 20899, USA}
 \email{{\red jamer@usna.edu}}
\date{\today}

\begin{abstract}
The effects of post-synthesis annealing temperature on arc-melted samples of Fe$_{3}$Ga$_{4}$ has been studied to investigate changes in crystallographic and magnetic properties induced by annealing. Results show a significant trend in the evolution of the (incommensurate spin density wave) ISDW-FM (ferromagnetic) transition temperature as a function of the refined unit cell volume in annealed samples. Strikingly, this trend allowed for the tuning of the transition temperature down to room-temperature (300 K) whilst maintaining a sharp transition in temperature, opening the door to the use of Fe$_{3}$Ga$_{4}$ in functional devices. Crystallographic analysis through Rietveld refinement of high-resolution x-ray diffraction data has showed that arc-melted stoichiometric Fe$_{3}$Ga$_{4}$ is multi-phase regardless of annealing temperature with a minor phase of FeGa$_{3}$ decreasing in phase fraction at higher annealing temperature. In order to validate the trend in ISDW-FM transition temperature with regard to unit cell volume, high pressure magnetometry was performed. This showed that the FM-ISDW ($\sim$ 68 K) and ISDW-FM ($\sim$ 360 K) transition temperatures could be tuned, increased and decreased respectively, linearly with external pressure. Thus, external pressure and the ensuing crystallographic changes minimize the temperature range of the stability of the ISDW pointing toward the importance of structural properties on the mechanism for the formation of the intermediate ISDW phase. These results show how this model system can be tuned as well as highlighting the need for future high-pressure crystallography and related single crystal measurements to understand the mechanism and nature of the intermediate ISDW phase to be exploited in future devices.
\end{abstract}

\maketitle

\section{Introduction}
There has been a push in recent years to find materials for use in the next-generation magnetic devices for spintronic and multicaloric applications. \cite{Hoffmann, Gschneidnerj, Jungwirth, Stern, Barua2} One class of interesting materials are antiferromagnets (AFM) with room temperature or near-room temperature magnetic transitions such as FeRh or metallic antiferromagnets. \cite{Manekar, Barua, Cress, Bennett, Roy, Yu} Metallic antiferromagnets allow for exotic interplay between spin, charge, and magnetization dynamics, which is a very exciting prospect for future devices. \cite{Hu, Siddiqui} Both avenues offer promise for useful innovations and a concerted effort toward material discovery is required to identify materials with ideal properties. To that end, one recently re-discovered material Fe$_{3}$Ga$_{4}$ exhibits both types of behavior which makes it a exemplary candidate for investigation from a fundamental point-of-view in order to understand the complex magnetic and physical properties as well as for possible use in multi-functional devices. \cite{Mendez}

Fe$_{3}$Ga$_{4}$ crystallizes in the $C2/m$ monoclinic space group with four unique Fe atoms which house a large amount of possibly significant nearest and next-nearest neighbor interactions which drive the magnetic properties in this system. \cite{Philippe} Previous studies on Fe$_{3}$Ga$_{4}$ have indicated that it displays multiple magnetic transitions from a ferromagnetic (FM) ground state to an intermediate AFM phase at 68 K to a re-entrant FM phase at 360 K, and a transition at 420 K to a paramagnetic (PM) state. \cite{Mendez, Al1995, Samatham, Kawamiya} The high temperature transition from AFM to FM at 360 K is of particular interest due to its proximity to room temperature as well as the unknown nature of the intermediate AFM phase. \cite{WuThesis, DuijnThesis} Recent work has shown evidence that the FM ground state and re-entrant FM state between 360 K and 420 K are the same state which means the intermediate AFM state is formed through a instability in the FM ground state. This has been supported by neutron diffraction and first-principles calculation which have shown evidence that the true nature of the intermediate AFM phase is an incommensurate spin density (ISDW) wave which is formed through a Fermi surface instability of the FM state. \cite{Wu} Unlike other examples where an ISDW forms from a FM state,\cite{Bhattacharyya, Abdul, Niklowitz} the ISDW state in Fe$_{3}$Ga$_{4}$ has a large temperature range of stability which makes it an ideal platform to study this exotic magnetic state. There are example of similar behavior in $f$-electron systems with a re-entrant FM state following an intermediate AFM state, \cite{Kumar, Wangf, Tomka} but Fe$_{3}$Ga$_{4}$ is the only example of an only $d$-electron system.

\begin{figure*}[ht!]
    \centering
    \includegraphics[width = 6.69in]{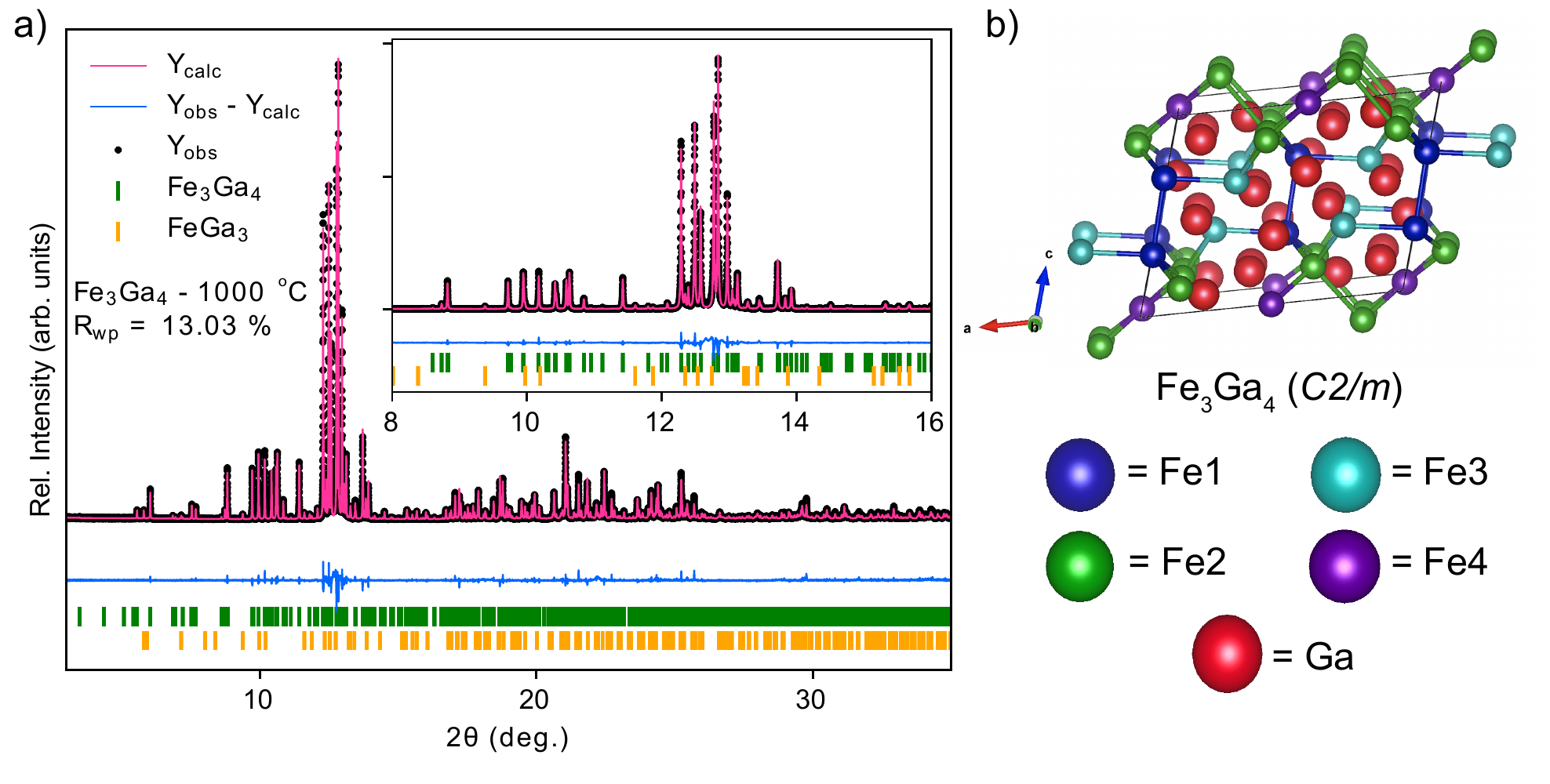}
    \caption{a) High-resolution synchrotron powder XRD pattern of Fe$_{3}$Ga$_{4}$ annealed at 1000 $^{\circ}$C for 48 hours at room temperature (about 295~K). Tick marks representing the corresponding binary phases are shown below the calculated, observed, and difference curves from Rietveld analysis. Fit statistics are summarized in the figure. b) Crystal structure of monoclinic Fe$_{3}$Ga$_{4}$ with the four unique Fe atoms distinguished with different colors and Fe-Fe bonds under 3 \AA~shown to emphasize nearest-neighbor Fe interactions. } 
    \label{fig:xrd}
\end{figure*}

This intermediate phase has been explored through pressure, doping, and high magnetic fields in order to better understand its origin and related properties. It has been shown that doping can effectively tune the IDSW to FM transition temperature as well as the low temperature FM to ISDW transition. \cite{Al1995, Al2000, Duijn, Kobeissi, Al1998}  Similarly, previous work has shown that external pressure is effective at tuning the low temperature FM-ISDW transition temperature, but the effect on the high temperature ISDW-FM transition temperature has not been reported. \cite{DuijnThesis, DujinPressure} From these works, it appears that intermediate IDSW is sensitive to effects caused by doping through electron count changes and/or structural changes which in turn effect the electronic structure in this regime; although this has not be rigorously studied. The transition from FM to ISDW is accompanied by a very small volume change but no structural transition is observed which further supports the SDW instability hypothesis. \cite{DujinPressure, DuijnThesis, Benavides} Thus, the observed effects of pressure on the magnetic properties are likely due to electronic structure changes concomitant with applied pressure. 

The object of this study is to look at both the effects of pressure and annealing temperature to determine if these can be used to tune the transition temperatures of Fe$_{3}$Ga$_{4}$ effectively. Previous work has stated that annealing temperature does effect the magnetic properties without substantial evidence to understand what drives these changes. \cite{Mendez} In this study, polycrystalline ingots of Fe$_{3}$Ga$_{4}$ were annealed at various temperatures, which lead to a controllable trend in the approximate composition of the Fe$_{3}$Ga$_{4}$ with respect to the known secondary phase, FeGa$_{3}$ which is non-magnetic and semiconducting. \cite{Gippius, Hadano, Haeussermann} The observation and quantification of the impurity FeGa$_{3}$ phase in polycrystalline arc-melted Fe$_{3}$Ga$_{4}$ has not been reported previously. We have used high-resolution synchrotron x-ray diffraction to determine the changes caused by annealing temperature and formation of secondary phases which has previously not been quantified. Magnetic measurements on these annealed samples at ambient pressure and high pressure are combined to further elucidate the cause of magnetic property changes accompanying structural changes due to synthesis method and external pressure. Combined, this work has shown that the magnetic properties of Fe$_{3}$Ga$_{4}$ can be tuned with unit cell volume either through synthetic procedure or external pressure. This observable change allows for the exotic re-entrant AFM-FM magnetic transition to be tuned to room temperature for possible use in future devices.


\section{Experimental details}

\subsection{Synthesis procedure}
Bulk samples of Fe$_{3}$Ga$_{4}$ were synthesized via arc-melting of elements in corresponding stoichiometric ratios using an Edmund Buehler MAM-1 under ultra-high purity Ar atmosphere. The arc-melted ingots were loaded into quartz tubes and sealed under moderate vacuum ($< 10^{-2}$ Torr). The samples were then annealed at various temperatures of 550, 700, 800, 900, 925, 950, 975, and 1000 $^{\circ}$C for 48 hours allowed to cool to room temperature naturally. 

\subsection{X-ray diffraction and elemental analysis}
High resolution synchrotron X-ray diffraction was performed on powders of ground as-recovered ingots at Beamline 11-BM at the Advanced Photon Source at Argonne National Lab. Ground powders were packed in 0.4 mm Kapton capillary tubes and sealed with clay under atmospheric conditions. Diffraction data was collected between 0.5 and 46 degrees with a step size of 0.0001$^{\circ}$ using a constant wavelength of 0.457921 \AA~at 300 K. Rietveld refinements and data analysis was performed using the GSAS software suite. \cite{GSAS} EDS spectra were collected on a LEO SEM/EDAX system at a beam energy of 20kV. The tabulated compositions were obtained as an averages of 5 different sites collected from a cleaved edge surface of each specimen.

\subsection{Magnetic measurements} 
A Quantum Design Dynacool Physical Property Measurement System (PPMS) was used to measure the magnetization of the Fe$_{3}$Ga$_{4}$ samples in magnetic fields up to 5 T and in the temperature interval of 300-800 K. Magnetic measurements under applied hydrostatic pressure (P) was performed in a commercial CuBe cylindrical pressure cell (Quantum Design). Daphne 7373 oil was used as the pressure transmitting medium. The applied pressure value was calibrated by measuring the shift of the superconducting transition temperature of Pb, which was used as a reference manometer. Errors due to differential thermal contraction between the pressure cell components and the pressure transmitting the medium was minimized by setting a 1 K/min temperature sweep-rate. Ambient pressure thermomagnetic characterization was carried out on all annealed samples using a Quantum Design SQUID MPMS system. Samples were first saturated in a field of 1500 Oe, then the field was removed and the sample was cooled to 5 K. A field of 100 Oe was then applied and the magnetic moment was measured while heating to 400 K to obtain the zero-field-cooled (ZFC) curve. The sample was then cooled back to 5 K and the magnetic moment was then measured again in a 100 Oe field while heating to 400 K to obtained the field-cooled curve (FC).


\begin{table*}[]
\begin{tabular}{|c|c|c|c|c|c|c|c|c|c|c|c|}
\hline
Temp. ($^{\circ}$C) & $a$ (\AA) & $b$ (\AA) & $c$  (\AA) & $\beta$ ($^{\circ}$)& Volume (\AA$^{3}$) & \% Fe$_{3}$Ga$_{4}$ & \% FeGa$_{3}$ & Calc. Stoich. & T$_{1}$ (K) & T$_{2}$ (K)  \\ \hline
550                       & 10.1023(1)           & 7.6692(2)             & 7.8750(5)              & 106.29(1)          & 585.66(1)                       & 86.8(1)            & 13.20(3)      & Fe$_{3}$Ga$_{3.99(2)}$ & 67 & 387      \\ \hline
700                       & 10.1013(1)            & 7.6694(2)             & 7.8755(5)              & 106.28(2)          & 585.66(2)                       & 83.1(1)             & 16.84(3)      & Fe$_{3}$Ga$_{3.94(2)}$  & 53 & 385      \\ \hline
800                       & 10.0971(2)           & 7.6719(2)             & 7.8771(5)              & 106.26(1)          & 585.77(1)                       & 82.3(1)             & 17.61(3)      & Fe$_{3}$Ga$_{3.92(2)}$  & 65 & 377          \\ \hline
900                       & 10.0982(3)            & 7.6881(2)             & 7.8726(5)              & 106.29(1)          & 585.18(1)                       & 92.1(1)             & 7.87(3)       & Fe$_{3}$Ga$_{3.95(2)}$  &  52 & 299       \\ \hline
925                       & 10.0997(1)            & 7.6692(1)             & 7.8756(3)              & 106.28(2)          & 585.56(1)                       & 90.1(1)             & 9.89(3)       & Fe$_{3}$Ga$_{3.96(2)}$  & 61 & 379       \\ \hline
950                       & 10.1000(2)            & 7.6687(1)             & 7.8751(4)              & 106.28(1)          & 585.50(1)                       & 93.5(2)             & 6.47(3)       & Fe$_{3}$Ga$_{3.97(2)}$  & 56 & 368        \\ \hline
975                       & 10.0986(1)            & 7.6685(1)             & 7.8751(2)              & 106.28(1)          & 585.38(1)                       & 94.3(2)             & 5.69(3)       & Fe$_{3}$Ga$_{3.95(2)}$   & 57 & 320      \\ \hline
1000                      & 10.0988(1)            & 7.6689(2)             & 7.8736(1)              & 106.29(1)          & 585.30(1)                       & 96.8(2)             & 3.11(2)       & Fe$_{3}$Ga$_{3.98(2)}$ & 55 & 318  \\ \hline         
\end{tabular}
\caption{Compositional analysis of arc-melted and annealed Fe$_{3}$Ga$_{4}$ from Rietveld refinement of synchrotron powder XRD data. The first five columns display the structural parameters for Fe$_{3}$Ga$_{4}$ in each annealing temperature, the next two columns show the phase fraction of Fe$_{3}$Ga$_{4}$ and FeGa$_{3}$ in each sample, and the final two columns show the calculated stoichiometry of Fe$_{3}$Ga$_{4}$ from Rietveld refinement of Fe/Ga occupancy and EDS elemental analysis results. All errors shown in parentheses for corresponding quantities.}
\label{table:lattice}
\end{table*}

\section{Results and Discussion}

\subsection{Crystallographic and compositional analysis}
Fe$_{3}$Ga$_{4}$ crystallizes in the monoclinic \textit{C2/m} space group with a complex network of four unique Fe atoms within the unit cell. These unique Fe atoms are shown in Figure \ref{fig:xrd} with a representative XRD powder pattern and refinement for the 1000 $^{\circ}$C annealed Fe$_{3}$Ga$_{4}$ sample. The XRD and refinement show that no unknown phases are present. Previous reports have stated that crystallographic changes caused by annealing, vacancy defects, or anti-site defects can cause changes in the magnetic order of this system as these changes alter the Fe atom network within the unit cell, but a systematic study has not been reported. \cite{Mendez} We have synthesized polycrystalline ingots of Fe$_{3}$Ga$_{4}$ through arc-melting of stoichiometric elements and subsequently annealed under vaccuum conditions at various temperature for 48 hours. High-resolution synchrotron x-ray diffraction (XRD) was employed in order to understand the crystallographic changes induced by annealing temperature and relate those to changes in the magnetic ordering of the system. Table \ref{table:lattice} shows the results of Rietveld refinement on all annealed Fe$_{3}$Ga$_{4}$ samples. From this data, we can observe that the secondary impurity phase FeGa$_{3}$ appears at all annealing temperature but the phase fraction is significantly decreased with annealing temperatures above 900 $^{\circ}$C, but is always present. This is expected as Fe$_{3}$Ga$_{4}$ does not have a congruent melting point within the Fe-Ga phase diagram. This is important to note as the synthesis method performed for this work is in line with previous polycrystalline work and this quantitative phase analysis of impurity FeGa$_{3}$ has not been performed in previous work.\cite{Al1995, Samatham, Kawamiya} The phase fraction of Fe$_{3}$Ga$_{4}$ begins to significantly increase above 900 $^{\circ}$C due to the melting point of FeGa$_{3}$ and Fe$_{3}$Ga$_{4}$ occurring at 824 and 906 $^{\circ}$C respectively. Additionally for stoichiometric mixture of Fe:Ga in 3:4 ratio, the liquidus line is accessible at temperatures above $\sim 950 ^{\circ}$C and this decreases with higher Ga ratio in the system. FeGa$_{3}$ is well-reported as a non-magnetic semiconductor\cite{Gippius, Hadano, Haeussermann} and thus its phase fraction should have no direct impact on the magnetic signal for these ingots. 

It would also appear that the phase fraction of FeGa$_{3}$ does not have a systematic impact on the stoichiometry of the Fe$_{3}$Ga$_{4}$ phase. This is determined from XRD refinement which is an average structure analysis - microstructure analysis could be important to understand how intergrown FeGa$_{3}$ grains affect the magnetic properties of Fe$_{3}$Ga$_{4}$ polycrystalline material. The refinement process allowed for Fe atom vacancies, Ga atom vacancies, and anti-site mixing after all structural and instrumental parameters have been refined to their optimum values. This process showed that forced Fe vacancies and anti-site mixing always negatively impacted the fit statistics and fit profile; however, Ga vacancies on the order of < 3 \% on some sites improved the refinement statistics. This makes sense as the starting material was stoichiometric and any non-zero amount of FeGa$_{3}$ impurity should affect the Ga stoichiometry of the Fe$_{3}$Ga$_{4}$ phase. The only other structural parameter which shows change across these annealing temperature is the unit cell volume which changes on the order of 0.1\% between these samples. However, the driving force for unit cell changes cannot be determined as its volume does not systematically trend with Ga vacancies or annealing temperature. Attempts to refine other crystallographic factors such as crystallite strain for example were not useful with the current data due to the multiphase nature of the samples. Future work on single phase Fe$_{3}$Ga$_{4}$ which has been post-annealed at various temperatures could offer insight into the affects of crystallite strain on the magnetic properties of the system. Lastly, the four unique Fe atoms in the monoclinic unit cell host a large number of nearest and next-nearest-neighbor interactions which all may play a critical role in the manifestation of magnetic order in this system. Rietveld analysis allows Fe-Fe bond distances to be determined and they are reported in Supplementary Information; however, they show no direct trends with annealing temperature.

\begin{figure*}[t]
    \centering
    \includegraphics[]{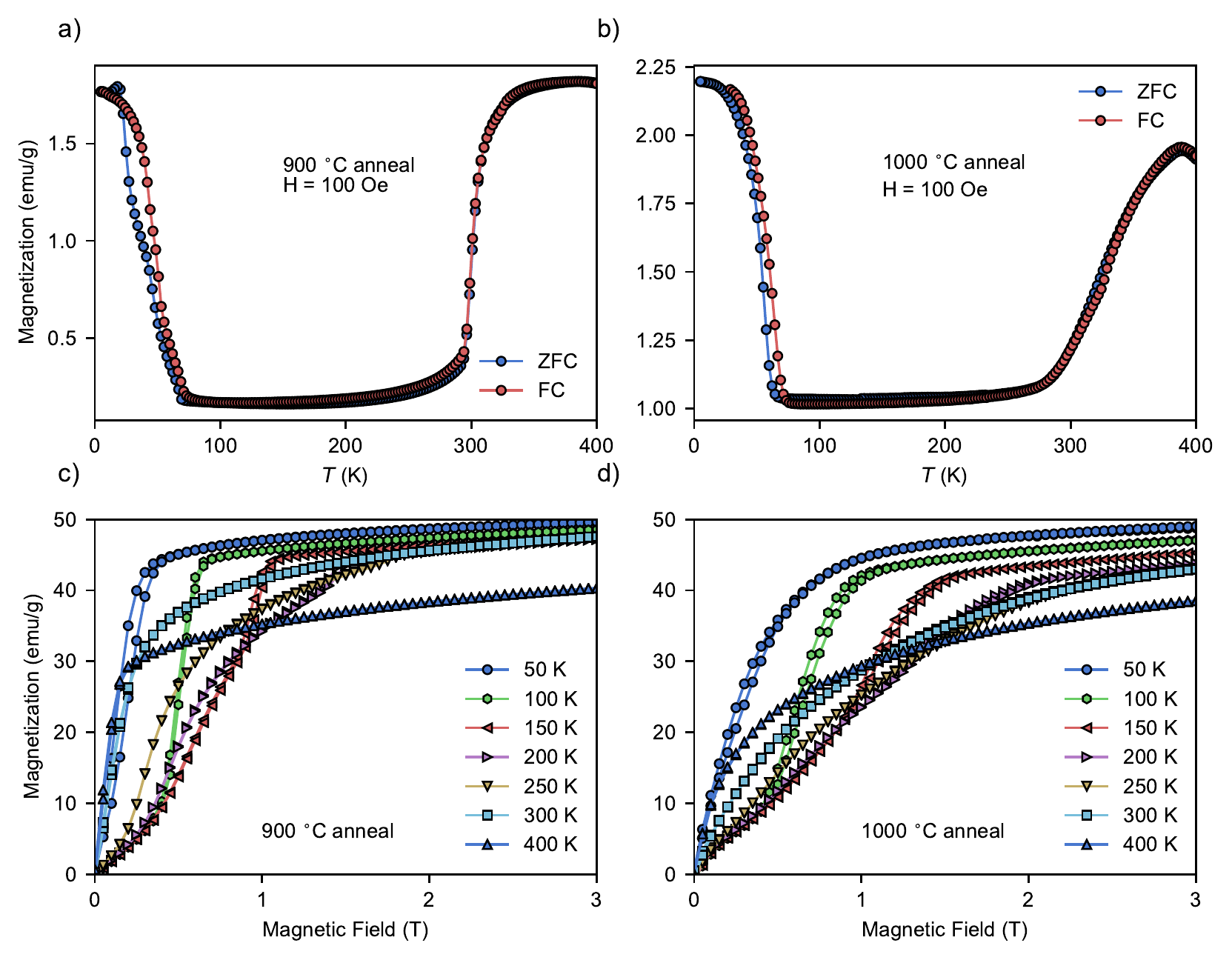}
    \caption{Temperature-dependent magnetization of Fe$_{3}$Ga$_{4}$ samples annealed at 900 $^{\circ}$C (a) and 1000 $^{\circ}$C (b) from 400 K - 1.8 K with an 100 Oe applied external field. Field-dependent magnetization of Fe$_{3}$Ga$_{4}$ samples annealed at 900 $^{\circ}$C (c) and 1000 $^{\circ}$C (d) from 0 - 3 T at various temperatures.} 
    \label{fig:mag}
\end{figure*}

\subsection{Ambient pressure magnetometry}
We have measured the temperature and field-dependent magnetization for all annealed samples and the data can be found in Supplementary Information with summarized transition temperatures in Table \ref{table:lattice}. For the remainder of this manuscript, we focus on the two most important samples which catalogue the magnetic properties of Fe$_{3}$Ga$_{4}$ at two different annealing temperature, 900 and 1000 $^{\circ}$C. These samples show significant differences in magnetic behavior. Figure \ref{fig:mag} shows the temperature-dependent magnetization at 100 Oe applied external field from 400 K to base temperature in the 900 and 1000 $^{\circ}$C annealed samples. The differences in the magnetic response of the two samples are clear - both samples have the same general behavior with a ferromagnetic (FM) ground state transitioning to an intermediate ISDW phase before returning to a re-entrant FM phase at high temperature. However, the ordering temperatures of the two are much different. The 1000 $^{\circ}$C sample shows expected behavior with a FM to ISDW transition around 68 K and an ISDW to FM transition in the range of previous work. \cite{Mendez, Al1995} The 900 $^{\circ}$C shows similar behavior for the FM-ISDW transition with a transition temperature of about 68 K, but the ISDW-FM transition is suppressed down to about 299 K and this is accompanied by a much sharper increase in magnetization upon heating through this range compared to the 1000 $^{\circ}$C sample. 

The change in ISDW-FM transition temperature between the 900 and 1000 $^{\circ}$C samples points toward a significant change in the energy of the ISDW phase relative to the FM phase. Figure \ref{fig:mag} shows the field-dependent magnetization at various temperatures for the 900 and 1000 $^{\circ}$C samples which further illustrates the differences between the two samples especially in the ISDW-FM transition temperature range. Previous work has shown that within the ISDW phase, a metamagnetic transition is observed with applied field and the required field to induce this metamagentic transition increases with temperature. \cite{Samatham, Mendez} Our data supports this observation for both samples. The key difference between the two samples is the response to field prior to the metamagnetic transition. At all temperatures within the ISDW phase prior to the metamagnetic transition, the change in magnetization with respect to applied field is much sharper for the 900 $^{\circ}$C sample compared to the 1000 $^{\circ}$C sample and at similar temperatures the required metamagnetic field is smaller for the 900 $^{\circ}$C sample. This can be seen at the 150 K isotherm in which the 900 $^{\circ}$C sample displays a metamagnetic transition at 1 T whereas the 1000 $^{\circ}$C displays the same transition at 1.5 T applied field. Similar phenomena was observed in single crystals oriented with the external field applied in different crystallographic directions - where field applied perpendicular to the $c$-axis induced the metamagnetic transition with smaller applied field than with field parallel to the $c$-axis. \cite{Mendez} Crystallographic anisotropy is unlikely to be the cause of this deviation from normal behavior seen in the 900 $^{\circ}$C sample since these samples are not single crystals. Therefore, when considering the decrease in ISDW-FM transition temperature as well as changes in the magnetic field response within the ISDW phase for the 900 $^{\circ}$C sample, there must be some external factor driving these observed changes in magnetism within this system. 

From Table \ref{table:lattice}, both samples contain a small amount of FeGa$_{3}$ and are Ga deficient on the order of less than 2\%, which is similar for the other annealed samples. The magnetic behavior of the other annealed samples is shown in Supplementary Information and from these, it is clear there is no direct trend between annealing temperature, phase fraction of FeGa$_{3}$ impurity and the ISDW-FM and FM-ISDW magnetic transition temperatures. However, from the Rietveld refinements of the annealed samples we can observe that there is a direct trend between the volume of the Fe$_{3}$Ga$_{4}$ unit cell and the ISDW-FM transition; we have shown this trend in Supplementary Information (all transition temperatures were determined by the inflection point of the derivative of magnetization with respect to temperature in the transition temperature range). This is important as it shows that this transition temperature can be altered to room temperature to allow for the possibility of this ISDW-FM phase transition to be incorporated for use in devices. Importantly, previous work showed that volumetric changes in Fe$_{3}$Ga$_{4}$ from above 400 K to 2 K is less than 0.5\% and the change from across the ISDW-FM transition is less than 0.1\% such that the observed change in unit cell volume cannot be solely attributed to structural changes between the ISDW and FM states and no structural transition accompanies the ISDW-FM magnetic transition.\cite{DuijnThesis, Benavides}  Future work is necessary to determine the factors which drive these measured changes in the unit cell volume caused by annealing or changes in the synthesis method in this system.  So, although we cannot currently tune unit cell volume from a growth perspective through annealing or defects/occupancy, we can simulate unit cell volume changes through external pressure and determine the changes in magnetic order of the system, especially within the ISDW phase.

\begin{figure}[t]
    \centering
    \includegraphics[]{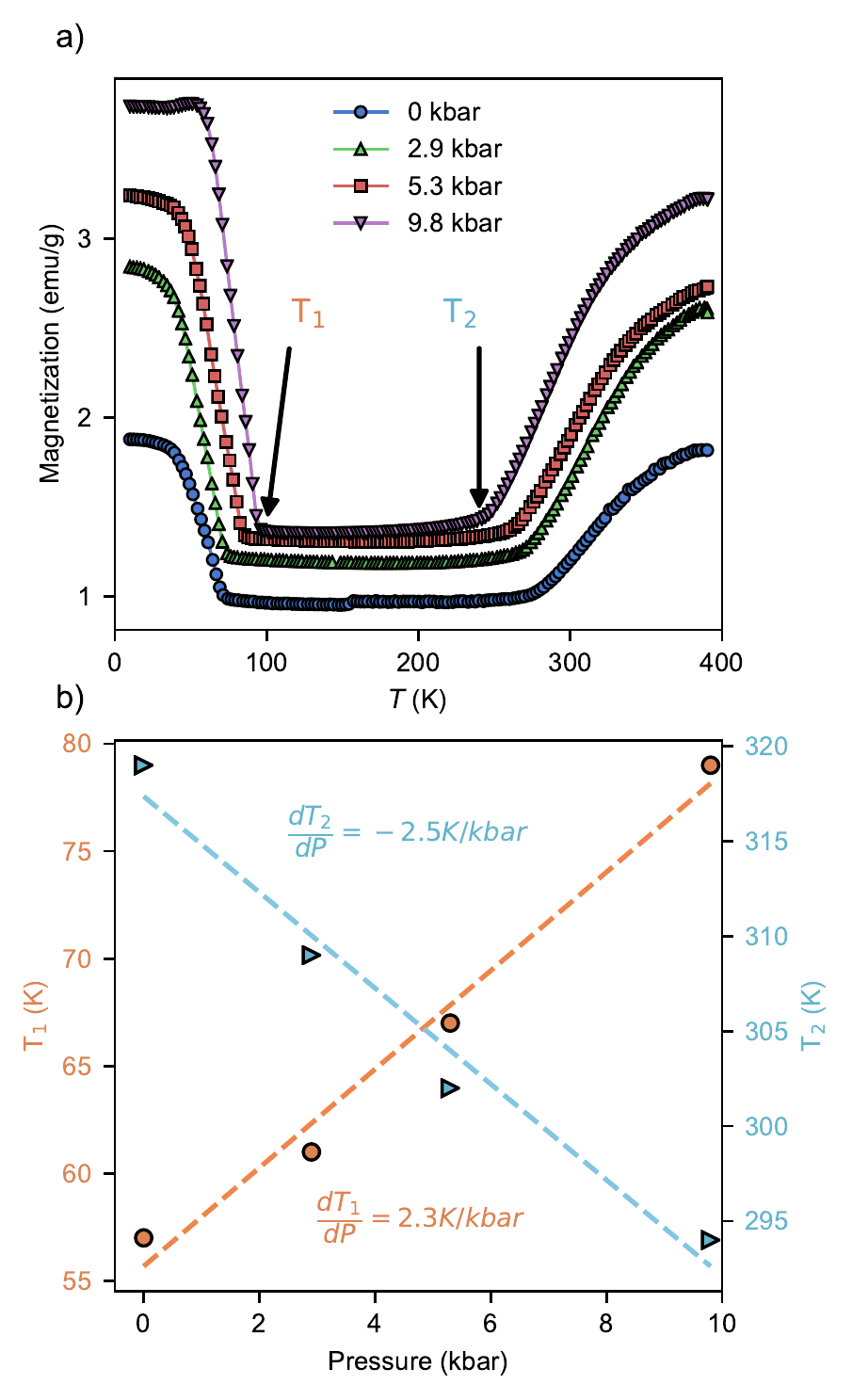}
    \caption{a) Temperature-dependent magnetization of 1000 $^{\circ}$C annealed sample from 400 - 1.8 K with an applied field of 100 Oe under various external pressures. b) Extracted transition temperature of the FM-ISDW (T$_{1}$), orange) transition and the ISDW-FM (T$_{2}$), teal) at the different applied external pressure. Transition temperatures were determined by the inflection point in derivative of magnetization with respect to temperature.}
    \label{fig:mag_pressure}
\end{figure}

\subsection{High-pressure magnetometry}
In order to reliably probe the affect of unit cell volume changes on the magnetic order of the system, we have measured the magnetic properties of the 1000 $^{\circ}$C sample under high pressure. Figure \ref{fig:mag_pressure} shows the magnetization as a function of temperature at various applied external pressure. We observe that external pressure tunes both the low temperature FM-ISDW transition as well as the high temperature ISDW-FM transition and these trends are summarized in Fig \ref{fig:mag_pressure}b for both transitions. In particular, the ISDW-FM transition is suppressed from around 320 K at ambient pressure down to below 295 K with 10 kbar of applied pressure which brings the transition temperature in line with the observed transition for the 900 $^{\circ}$C sample. Concurrently, the low temperature FM-ISDW transition temperature is increased meaning that the temperature regime for the stability of the ISDW phase is growing smaller with applied pressure. Combined, these two factors show that unit cell volume, which is directly being changed through external pressure and was inherently smaller in the 900 $^{\circ}$C sample, can be used to tune the intermediate ISDW phase in this system. This points toward a significant impact of crystallographic parameters on the stability of the ISDW phase with regard to its mechanism of formation and stabilization. Previous neutron work has suggested that the high temperature FM phase and low temperature FM phase are identical and the intermediate ISDW phase arises from an Fermi surface instability. \cite{Wu} This work showed that itinerant magnetism is co-existing with rather large localized magnetic moments of the Fe sites in the Fe$_{3}$Ga$_{4}$ lattice and thus a coupling mechanism between itinerant spins and the localized moments must play a role in stabilizing this magnetic order in this system. In previous systems, the spin density wave formation is driven by Fermi surface nesting by which different portions of the Fermi surface are coupled by the interaction of itinerant spins and local moments via the RKKY interaction. \cite{Feng, Overhauser, Overhauser2} If this type of mechanism holds for Fe$_{3}$Ga$_{4}$ then the effects of pressure and unit cell volume changes should affect the stability of the intermediate ISDW phase through electronic structure and/or magnetic interaction changes. Further work would be necessary to uncover the crystallographic changes caused by external high pressure to relate those changes to the changes in ISDW ordering temperature and stability. 

\begin{figure}[t]
    \centering
    \includegraphics[width = 3.66in]{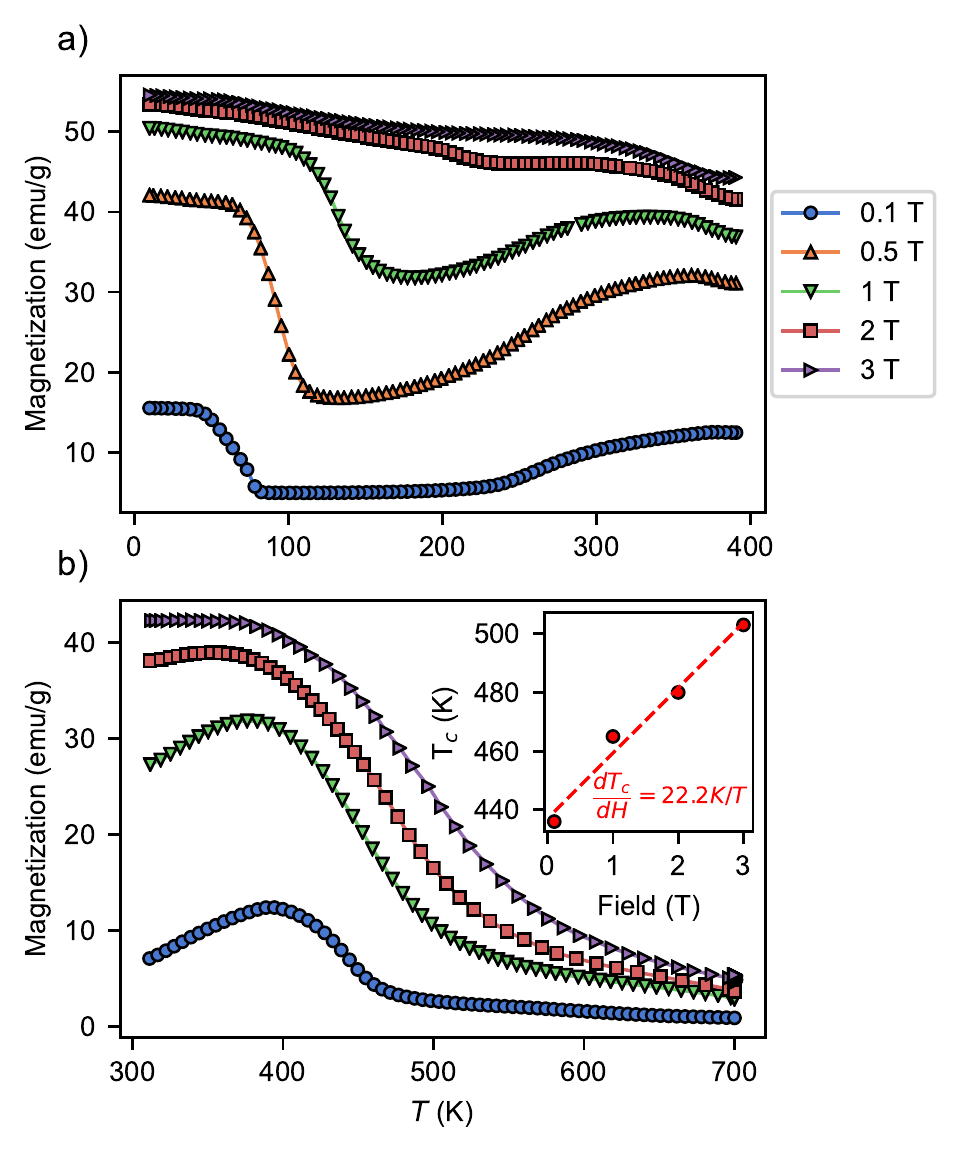}
    \caption{a) Temperature-dependent magnetization of 1000 $^{\circ}$C annealed sample from 400 - 1.8 K at various applied external fields showing the suppression of the intermediate ISDW at high applied fields. b) Temperature-dependent magnetization of 1000 $^{\circ}$C annealed sample from 400 - 1.8 K at various applied external fields with inset showing the change in FM-PM transition temperature as a function of applied field.} 
    \label{fig:mag_field}
\end{figure}

We have shown that the intermediate ISDW phase is sensitive to applied external pressure or internal pressure due to crystallographic factors and previous work has shown that this phase can be affected by applied field as well, but only below 400 K has been reported. \cite{Mendez, Samatham, Al1995} Figure \ref{fig:mag_field}a shows the temperature-dependent magnetization at various applied fields below 400 K. At 0.1 T applied field, the aforementioned low temperature FM-ISDW transition and high temperature ISDW-FM transition are clear and match well with previous reports. As the applied field is increased, the FM-ISDW transition temperature is increased in temperature with applied field and the ISDW-FM transition is broadened greatly with applied field. At high enough field, > 2 T, which is the maximum metamagnetic field required in the ISDW phase, the ISDW phase is seemingly fully suppressed so that the entire magnetization signal below 400 K is ferromagnetic owning from the FM-PM transition at 420 K above the measured range. From this, it appears that the intermediate ISDW can be fully suppressed to stabilize the ferromagnetic ground state across the entire temperature regime below the initial FM-PM transition with high applied field. Furthermore, we have measured the same properties in the range from 400 - 700 K in order to understand the effects of applied field for the FM-PM transition. Figure \ref{fig:mag_field}b shows the temperature-dependent magnetization at various applied fields from 300 - 700 K. At low fields (0.1 T), the ISDW-FM transition is clear as well as the FM-PM transition which occurs at 420 K, matching previous reports. \cite{Mendez} As expected, higher applied fields suppresses the ISDW-FM transition until it is no longer observable. Curiously, the FM-PM transition temperature increases with applied field as shown by Figure \ref{fig:mag_field}b inset. It is unclear what is driving this change but similar behavior has been noted in other systems and discussed as the observation of itinerant metamagnetism in the region of the ferromagnetic to paramagnetic transition. \cite{CoS1, CoS2, CoS3, levitin, Goto, Yoshimura} For this reason, this anomalous behavior in the region of the FM-PM transition requires further investigation to understand the field-dependence especially above the proposed ferromagnetic transition.  

\section{Conclusions}

In summary, we have studied the affects of variation in annealing temperature on crystallographic and magnetic properties of Fe$_{3}$Ga$_{4}$ through post-synthesis modification of arc-melted samples. High-resolution synchrotron XRD and subsequent Rietveld refinement was performed on all annealed samples to reveal variations in composition and structural parameters induced by annealing temperature. All annealed samples are multiphase containing Fe$_{3}$Ga$_{4}$ and a small phase fraction of non-magnetic semiconducting FeGa$_{3}$\cite{Gippius, Hadano, Haeussermann} where the phase fraction of the impurity phase could be suppressed to a minimum at 1000 $^{\circ}$C annealing temperature. Due to non-zero phase fraction of FeGa$_{3}$, all Fe$_{3}$Ga$_{4}$ phases were found to be Ga deficient on the order of less an 2\% without any observable anti-site mixing or Fe vacancies. Ambient pressure magnetic property measurements showed general behavior for Fe$_{3}$Ga$_{4}$ in agreement with previous work. \cite{Mendez, Al1995} In particular, previous work showed  metamagnetic transition to a field-polarized paramagnetic state in the ISDW phase and this metamagnetic field increased with temperature. Our work showed that at high enough fields, the ISDW-FM state could be completely suppressed yielding one observable FM-PM transition above 420 K that can also be altered by applied magnetic field. We have shown the ISDW-FM transition temperature is dependent on refined unit cell volume and can be altered from its reported value of 360 K to room temperature (299 K) upon a decrease in unit cell volume. In order to further probe this change, we have measured the magnetic properties of the 1000 $^{\circ}$C sample under high pressure to simulate the effect of systematic unit cell volume decrease. High pressure magnetometry revealed both the FM-ISDW and the ISDW-FM transition temperatures are pressure dependent where the FM-ISDW transition temperature increases with pressure and ISDW-FM transition temperature decreases. The net result of these effects is to decrease the temperature range of stability of the intermediate ISDW. This points toward the importance of crystallographic properties to the proposed mechanism for the existence of the ISDW phase in this compound. Future work on high pressure XRD to study the crystallographic changes under high pressure would be very useful to discern the structural changes concomitant with the changes in magnetic properties. Furthermore, single crystal growth and corresponding strain/pressure experiments could also help elucidate the mechanism of the ISDW phase and how it can be tuned for use in devices. \\

\section*{Supplemental Information}
See the supplementary material for the additional data. Additional data contains high-resolution XRD data for all Fe$_{3}$Ga$_{4}$ samples with accompanying Rietveld refinements with data summarized in Table I of the main text. Magnetometry data for all Fe$_{3}$Ga$_{4}$ samples is included as well, with temperature-dependent magnetization with magnetic transition temperatures extracted and field-dependent magnetization at various temperatures for all samples. Evolution of the FM-ISDW (T$_{1}$) and ISDW-FM  (T$_{2}$) transitions as a function of refined unit cell volume is included to show the linear change of T$_{2}$ with respect to unit cell volume. 

\section*{Data Availability}
The data that support the findings of this study are available from the corresponding author upon reasonable request.

\begin{acknowledgements}
Research at the United States Naval Academy was supported by the NSF DMR-EPM 1904446 and ONR 1400844839. Use of the Advanced Photon Source at Argonne National Laboratory was supported by the U.S. Department of Energy, Office of Science, Office of Basic Energy Sciences, under Contract No. DE-AC02-06CH11357. This work was supported by the U.S. Office of Naval Research through the Naval Research Laboratory's basic research program. Work at VCU was partially funded by National Science Foundation, Award Number: 1726617.
\end{acknowledgements}

\label{References}

\pagebreak
\setcounter{figure}{0}
\setcounter{table}{0}
\makeatletter 
\renewcommand{\thefigure}{S\@arabic\c@figure}
\makeatother
\makeatletter 
\renewcommand{\thetable}{S\@arabic\c@table}
\makeatother

\begin{center}

\textbf{Supplementary Material for\\ "Altering the magnetic ordering of Fe$_{3}$Ga$_{4}$ via thermal annealing and hydrostatic pressure"}

\end{center}

\textbf{Results of Rietveld powder X-ray diffraction analysis}
\vspace{5mm}

X-ray diffraction for the Fe$_{3}$Ga$_{4}$ samples annealed at various temperatures were taken at Beamline 11-BM at the Advanced Photon Source at Argonne National Lab between 0.5 and 46 degrees with a step size of 0.0001 using a constant wavelength of 0.457921 \AA~at 300 K. The sample was rotated continuously. The experimental data is (points) and fit (pink curve) are shown. The lower blue curve plots the experimental data minus the Rietveld fit. Tick marks in each figure represent the peak positions for each binary in the samples are determined from Rietveld analysis. The corresponding table (Table \ref{table:Fe}) catalogues Fe-Fe bond distances less than 3 \AA~in the Fe$_{3}$Ga$_{4}$ unit cell for all annealed samples as extracted from Rietveld refinement. 

\begin{table*}[t]
\begin{tabular}{|c|c|c|c|c|c|c|c|c|}
\hline
Temp. ($^{\circ}$C) & Fe$_{1}$-Fe$_{1}$ (\AA) & Fe$_{1}$-Fe$_{2}$ (\AA) & Fe$_{1}$-Fe$_{3}$ (\AA) & Fe$_{2}$-Fe$_{2}$ (\AA) & Fe$_{2}$-Fe$_{3}$ (\AA) & Fe$_{2}$-Fe$_{4}$ (\AA) \\ \hline
550                       & 2.9735  & 2.7716  & 2.7546  & 2.9300  & 2.5848  & 2.5598  \\ \hline
700                       & 2.9738  & 2.7718  & 2.7543  & 2.9302  & 2.5848  & 2.5597  \\ \hline
800                       & 2.9478  & 2.7710  & 2.7543  & 2.9080  & 2.6152  & 2.5625  \\ \hline
900                       & 2.9848  & 2.7607  & 2.7558  & 2.9080  & 2.5899  & 2.5655  \\ \hline
925                       & 2.9573  & 2.7716  & 2.7539  & 2.9224  & 2.5875  & 2.5667  \\ \hline
950                       & 2.9653  & 2.7636  & 2.7557  & 2.9293  & 2.5858  & 2.5760  \\ \hline
975                       & 2.9571  & 2.7677  & 2.7418  & 2.9031  & 2.6138  & 2.5639  \\ \hline
1000                      & 2.9534  & 2.7579  & 2.7552  & 2.9147  & 2.6003  & 2.5755  \\ \hline
\end{tabular}
\caption{Extracted bond distances from Rietveld refinement for annealed Fe$_{3}$Ga$_{4}$ samples between Fe atoms with less than 3 \AA~ distance. Labels for Fe atoms correspond to the positions noted in the unit cell depiction in the main text. FM-ISDW (T$_{1}$) and ISDW-FM (T$_{2}$) transition tempertures for each sample.}
\label{table:Fe}
\end{table*}

\begin{figure*}[]
    \centering
    \includegraphics[]{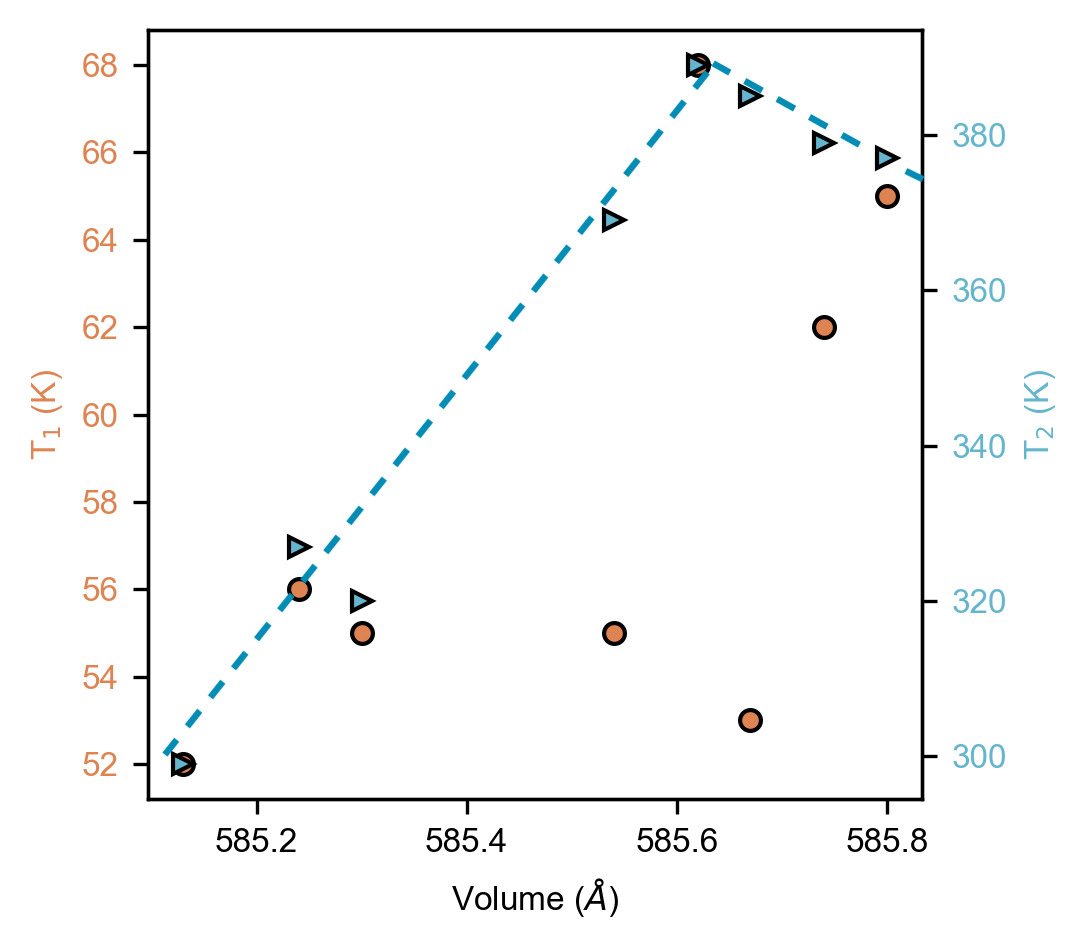}
    \caption{Evolution of FM-ISDW (T$_{1}$, orange) and ISDW-FM (T$_{2}$, blue) magnetic transition temperatures with respect to refined unit cell volume from Rietveld analysis of high-resolution powder XRD from 11-BM at Argonne National Laboratory. A direct trend is observed for the evolution of the ISDW-FM magnetic transition which is supported by accompanying high pressure magnetometry of the 1000 $^{\circ}$C sample presented in the main text. Dotted line as a guide to the eyes for the volume evolution of the T$_{2}$ ISDW-FM transition.} 
    \label{}
\end{figure*}

\begin{figure*}[]
    \centering
    \includegraphics[width = 6.69in]{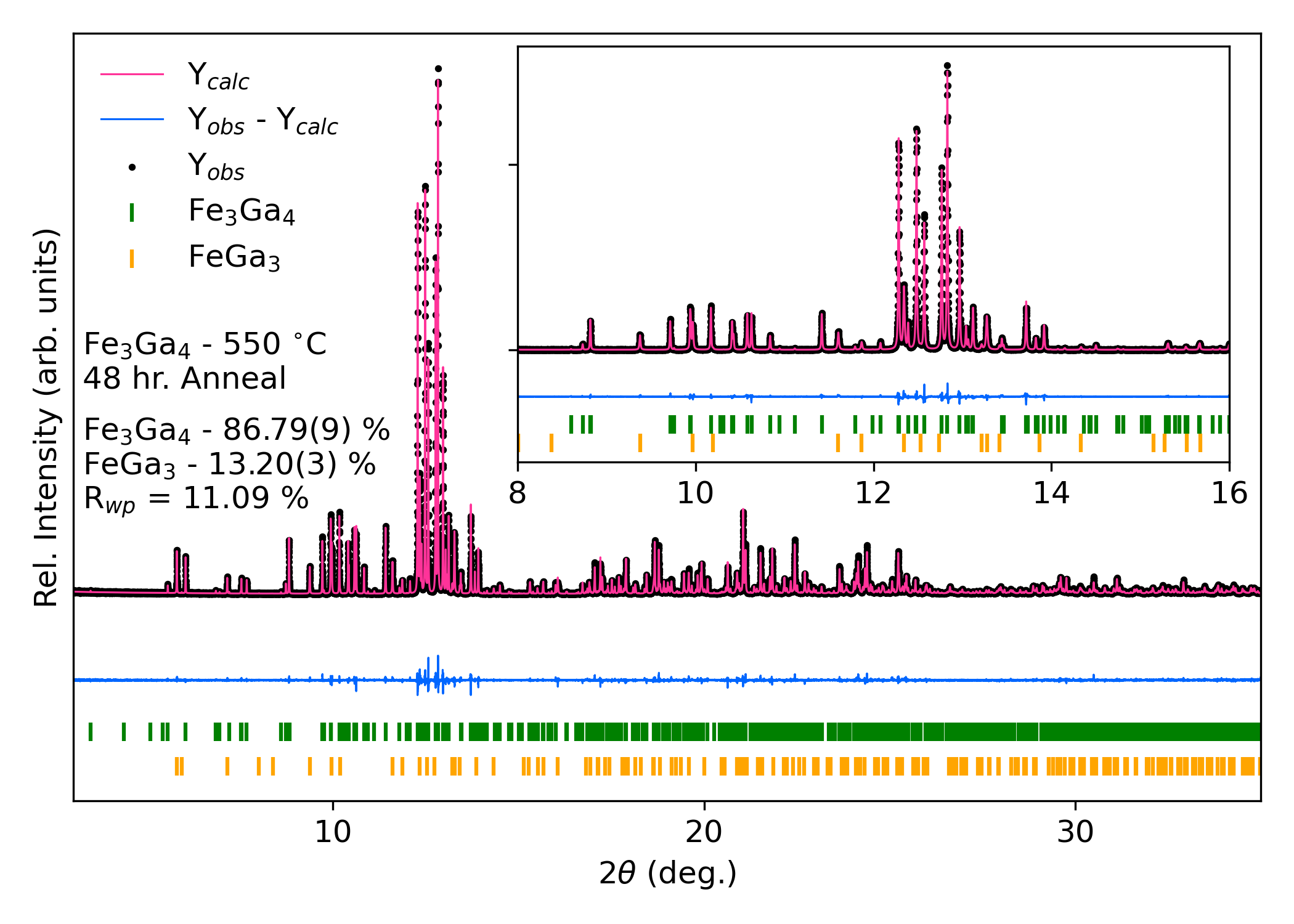}
    \caption{High-resolution synchrotron powder XRD pattern of Fe$_{3}$Ga$_{4}$ annealed at 550 $^{\circ}$C for 48 hours at room temperature (about 295~K). Tick marks representing the corresponding binary phases are shown below the calculated, observed, and difference curves from Rietveld analysis. Fit statistics are summarized in the figure} 
    \label{}
\end{figure*}

\begin{figure*}[]
    \centering
    \includegraphics[width = 6.69in]{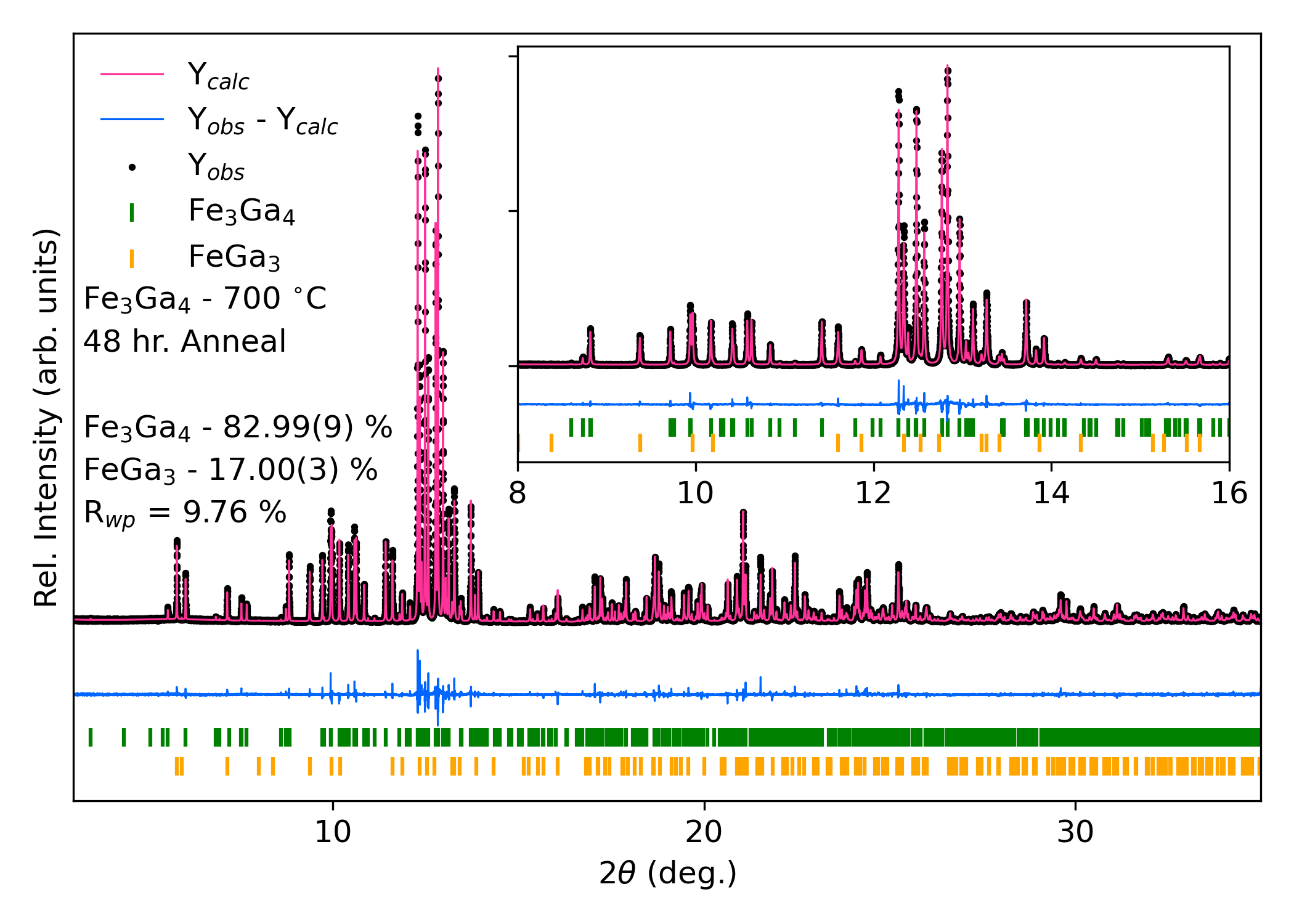}
    \caption{High-resolution synchrotron powder XRD pattern of Fe$_{3}$Ga$_{4}$ annealed at 700 $^{\circ}$C for 48 hours at room temperature (about 295~K). Tick marks representing the corresponding binary phases are shown below the calculated, observed, and difference curves from Rietveld analysis. Fit statistics are summarized in the figure} 
    \label{}
\end{figure*}

\begin{figure*}[]
    \centering
    \includegraphics[width = 6.69in]{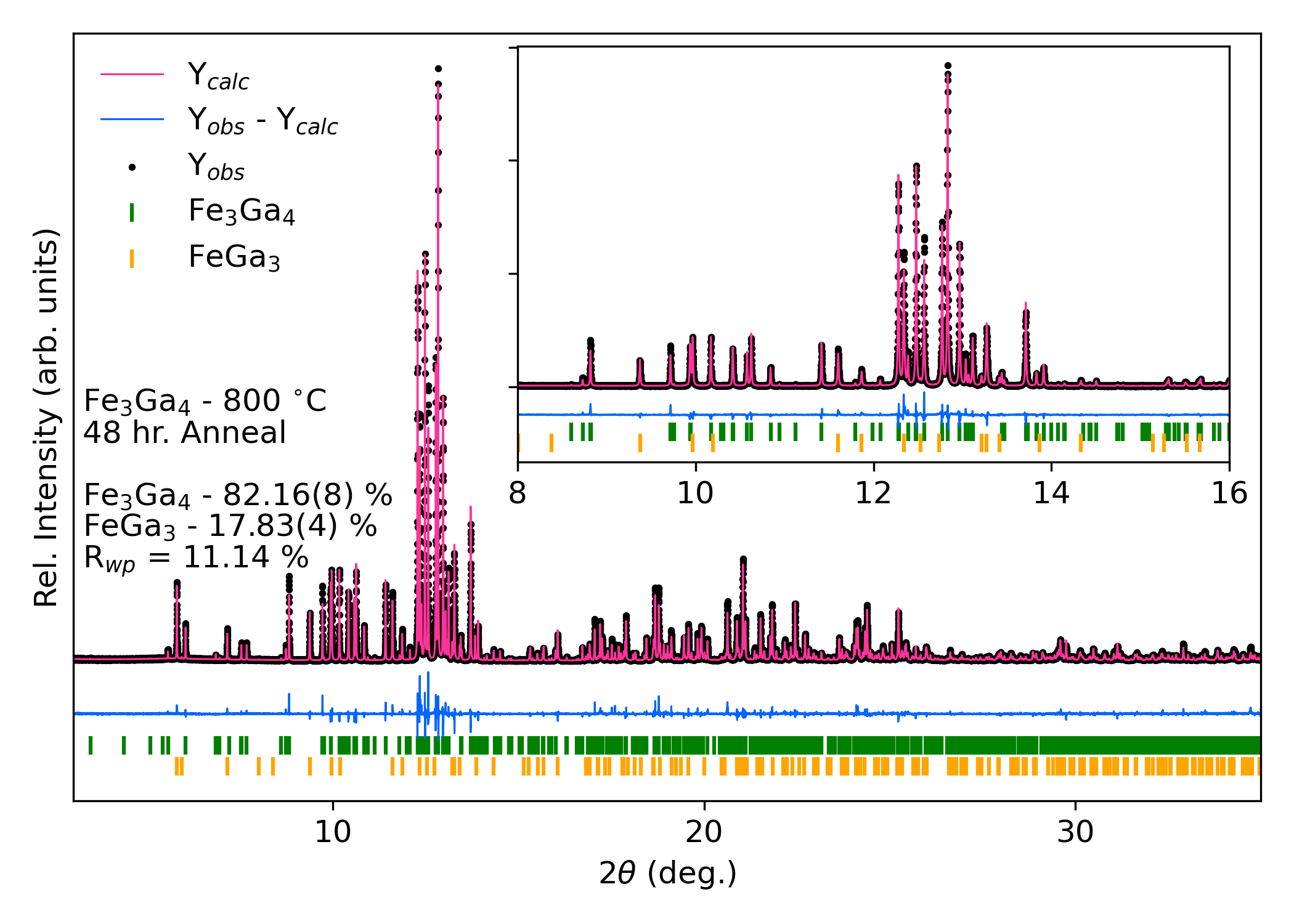}
    \caption{High-resolution synchrotron powder XRD pattern of Fe$_{3}$Ga$_{4}$ annealed at 800 $^{\circ}$C for 48 hours at room temperature (about 295~K). Tick marks representing the corresponding binary phases are shown below the calculated, observed, and difference curves from Rietveld analysis. Fit statistics are summarized in the figure} 
    \label{}
\end{figure*}

\begin{figure*}[]
    \centering
    \includegraphics[width = 6.69in]{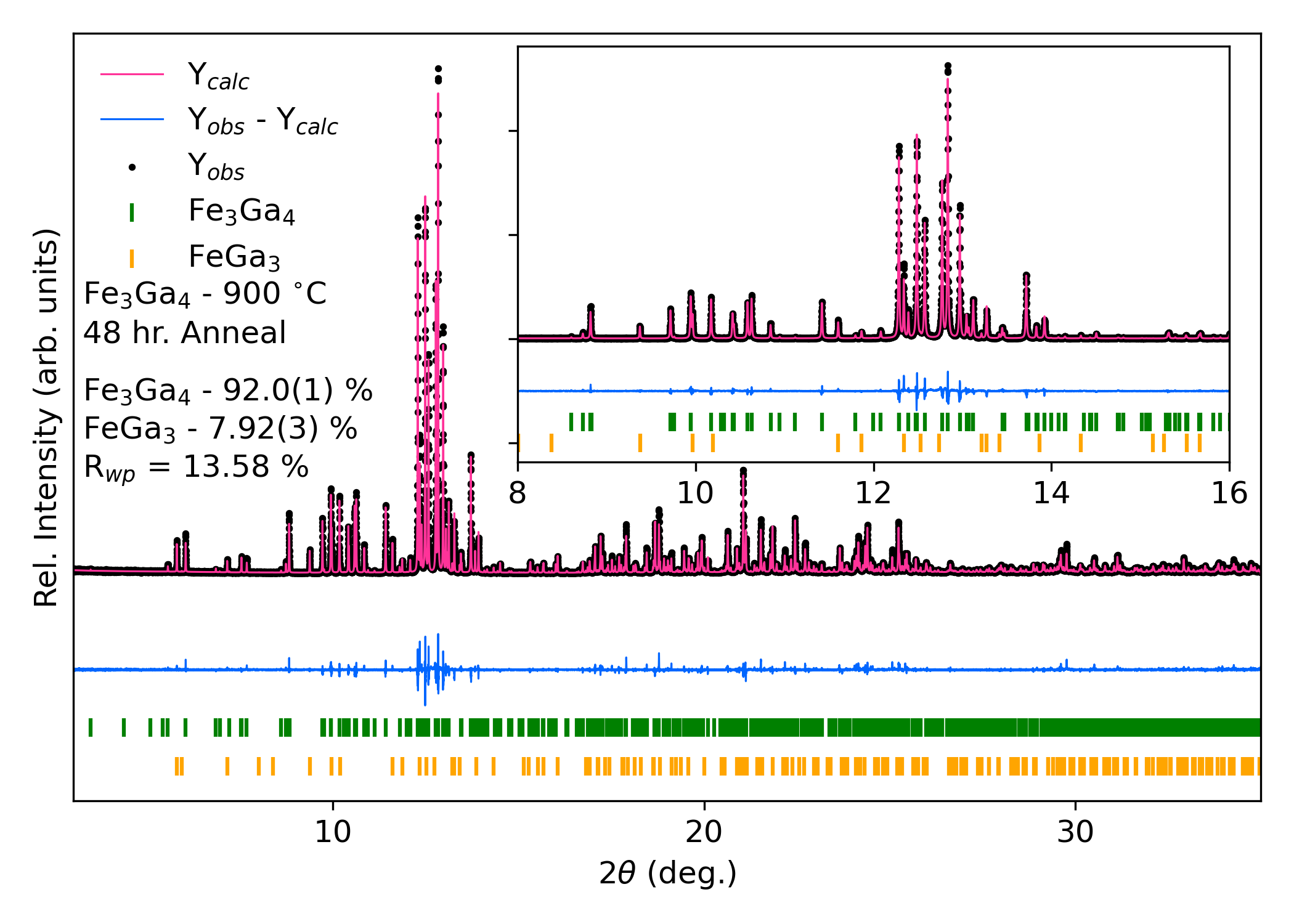}
    \caption{High-resolution synchrotron powder XRD pattern of Fe$_{3}$Ga$_{4}$ annealed at 900 $^{\circ}$C for 48 hours at room temperature (about 295~K). Tick marks representing the corresponding binary phases are shown below the calculated, observed, and difference curves from Rietveld analysis. Fit statistics are summarized in the figure} 
    \label{}
\end{figure*}

\begin{figure*}[]
    \centering
    \includegraphics[width = 6.69in]{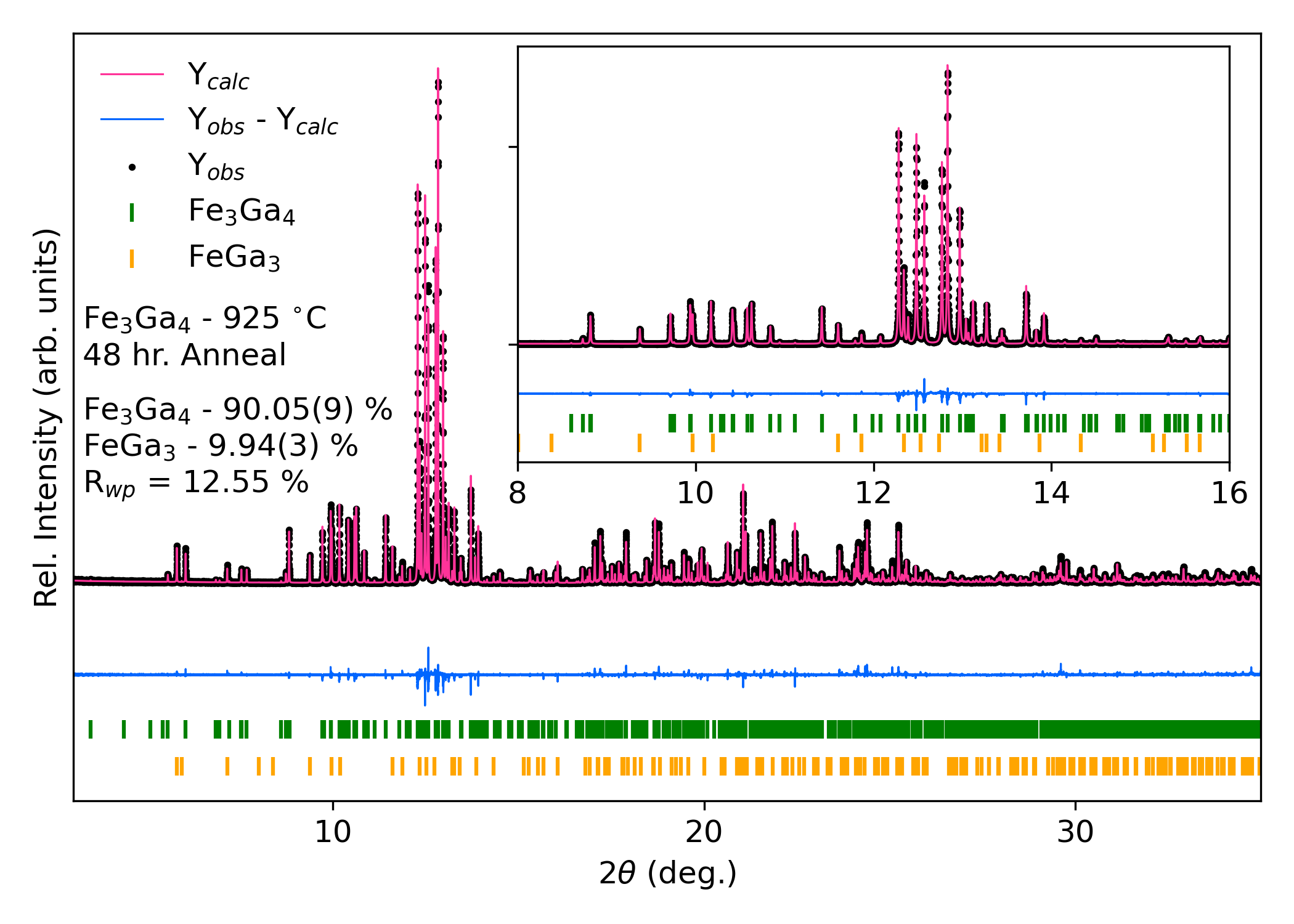}
    \caption{High-resolution synchrotron powder XRD pattern of Fe$_{3}$Ga$_{4}$ annealed at 925 $^{\circ}$C for 48 hours at room temperature (about 295~K). Tick marks representing the corresponding binary phases are shown below the calculated, observed, and difference curves from Rietveld analysis. Fit statistics are summarized in the figure} 
    \label{}
\end{figure*}

\begin{figure*}[]
    \centering
    \includegraphics[width = 6.69in]{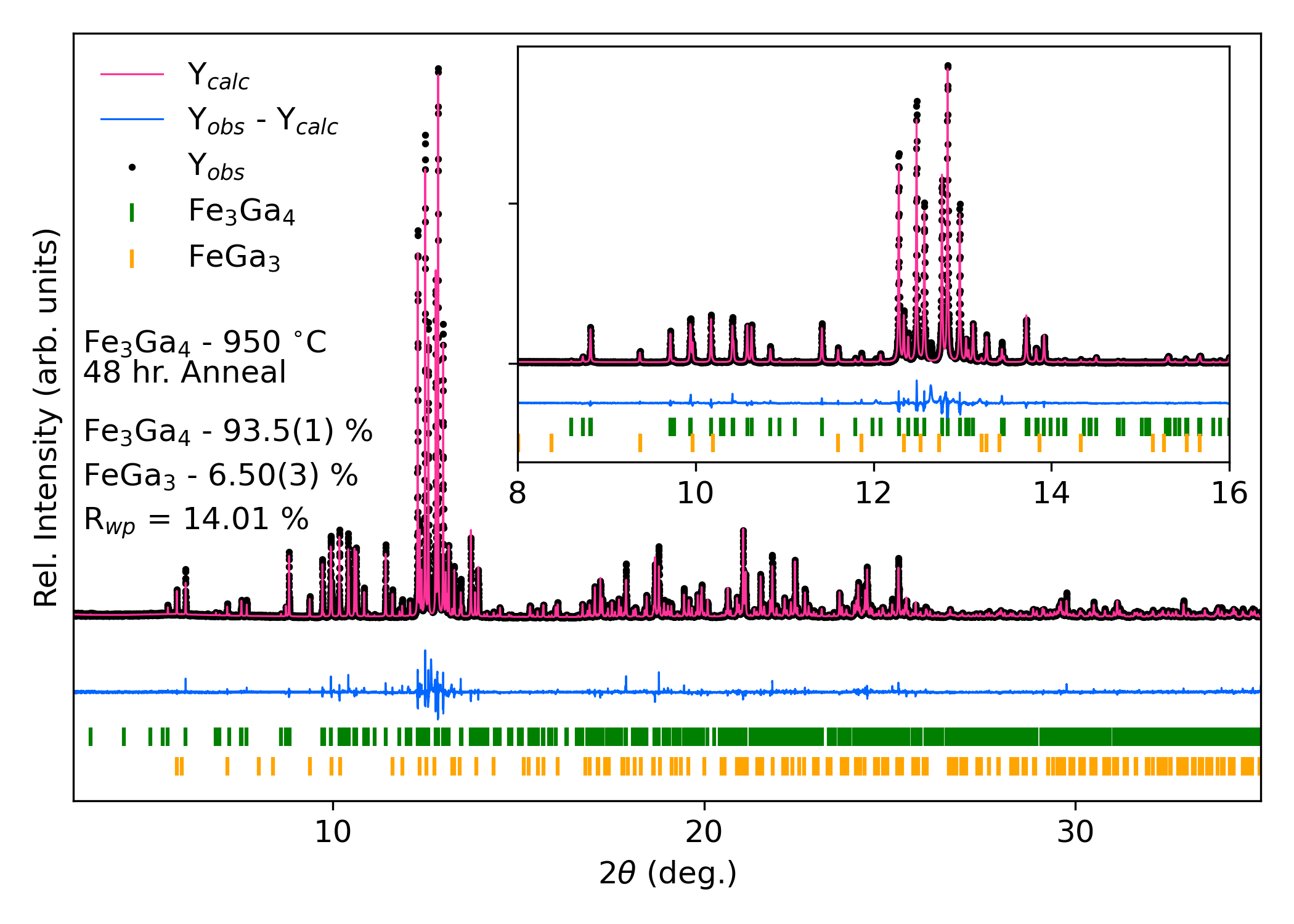}
    \caption{High-resolution synchrotron powder XRD pattern of Fe$_{3}$Ga$_{4}$ annealed at 950 $^{\circ}$C for 48 hours at room temperature (about 295~K). Tick marks representing the corresponding binary phases are shown below the calculated, observed, and difference curves from Rietveld analysis. Fit statistics are summarized in the figure} 
    \label{}
\end{figure*}

\begin{figure*}[]
    \centering
    \includegraphics[width = 6.69in]{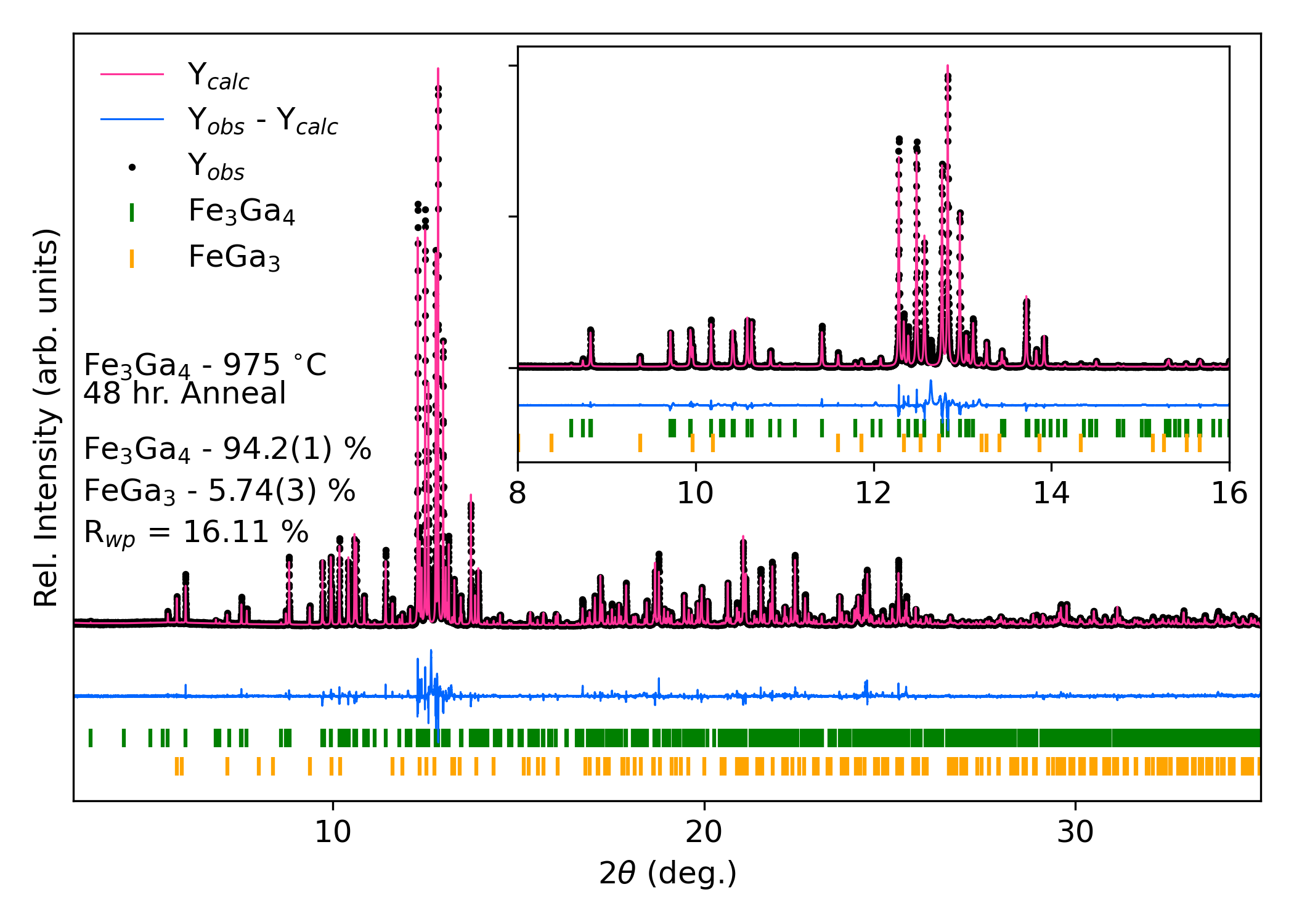}
    \caption{High-resolution synchrotron powder XRD pattern of Fe$_{3}$Ga$_{4}$ annealed at 975 $^{\circ}$C for 48 hours at room temperature (about 295~K). Tick marks representing the corresponding binary phases are shown below the calculated, observed, and difference curves from Rietveld analysis. Fit statistics are summarized in the figure} 
    \label{}
\end{figure*}

\begin{figure*}[]
    \centering
    \includegraphics[width = 6.69in]{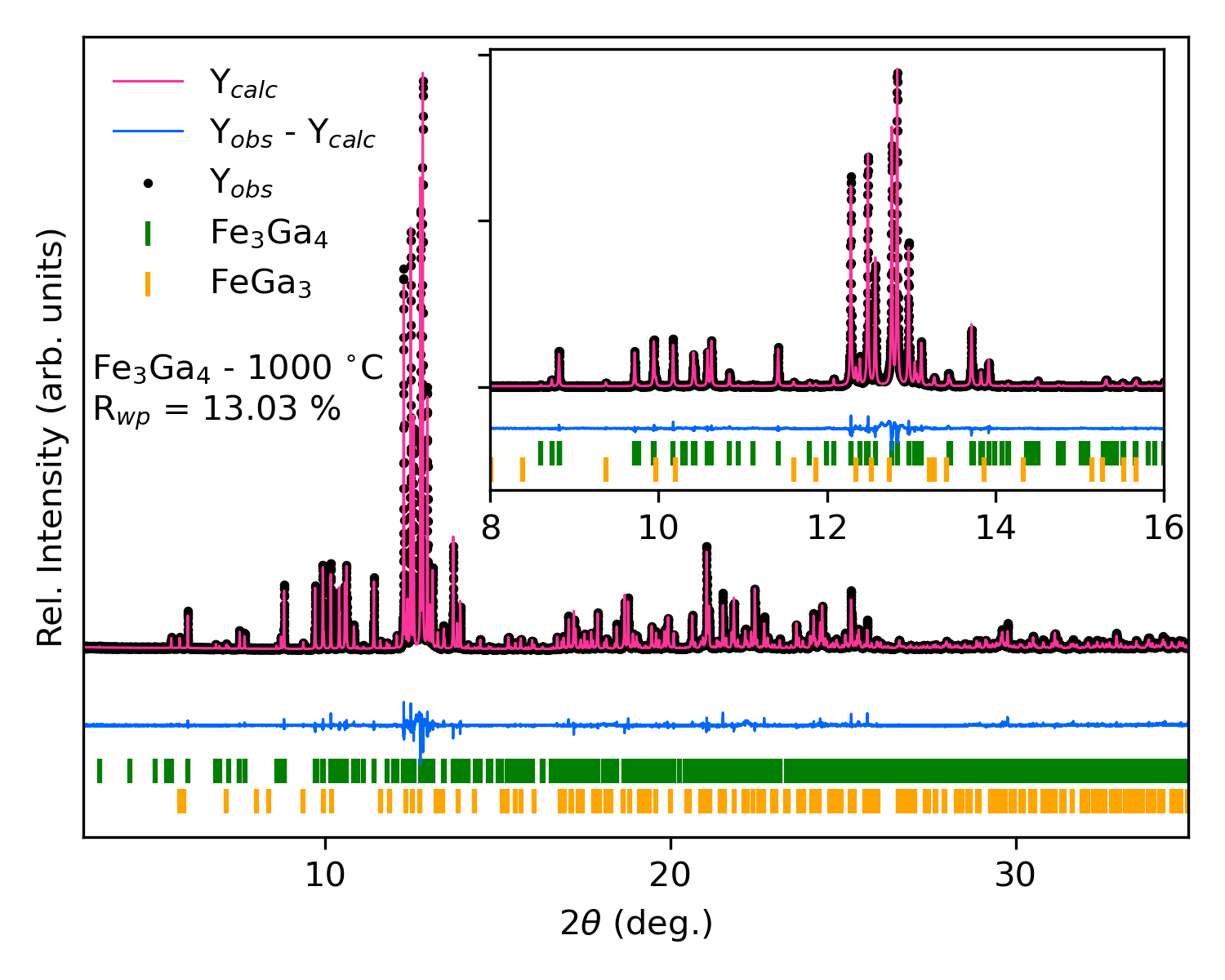}
    \caption{High-resolution synchrotron powder XRD pattern of Fe$_{3}$Ga$_{4}$ annealed at 1000 $^{\circ}$C for 48 hours at room temperature (about 295~K). Tick marks representing the corresponding binary phases are shown below the calculated, observed, and difference curves from Rietveld analysis. Fit statistics are summarized in the figure} 
    \label{}
\end{figure*}

\begin{figure*}[]
    \centering
    \includegraphics[height = 3.2in, width = 3.2in]{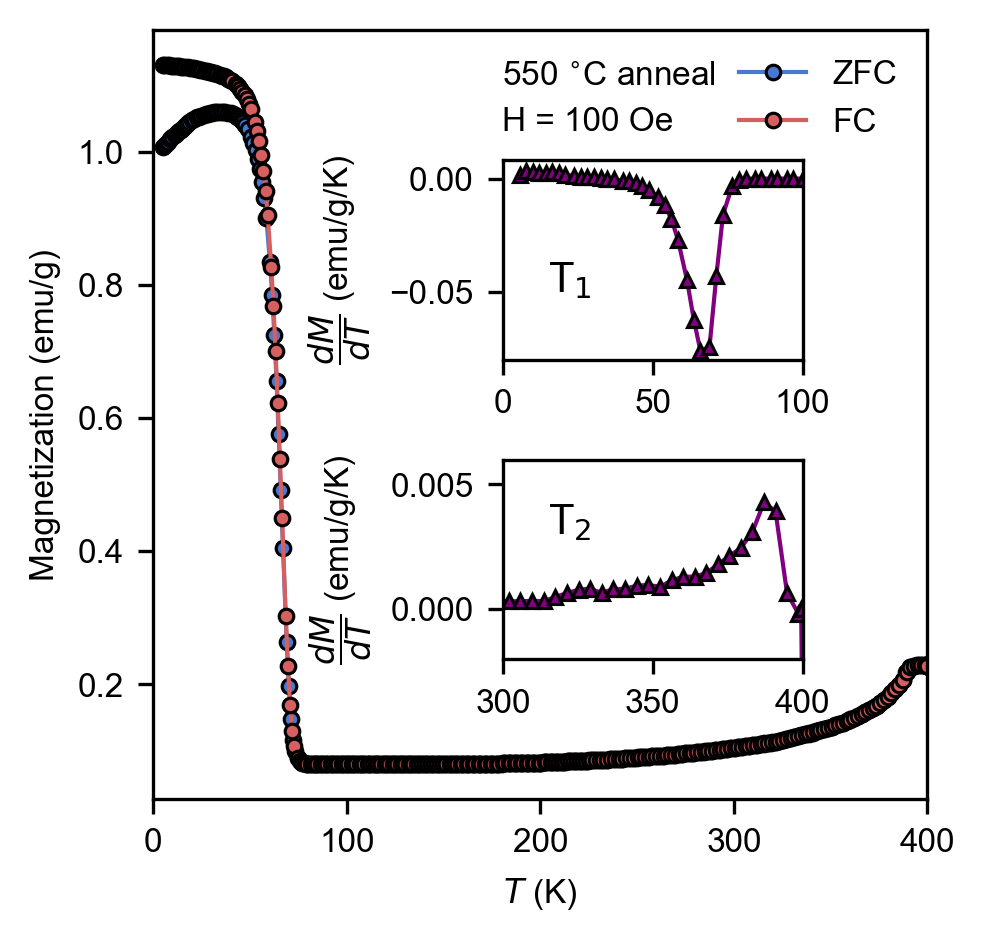}
    \includegraphics[height = 3.2in, width = 3.2in]{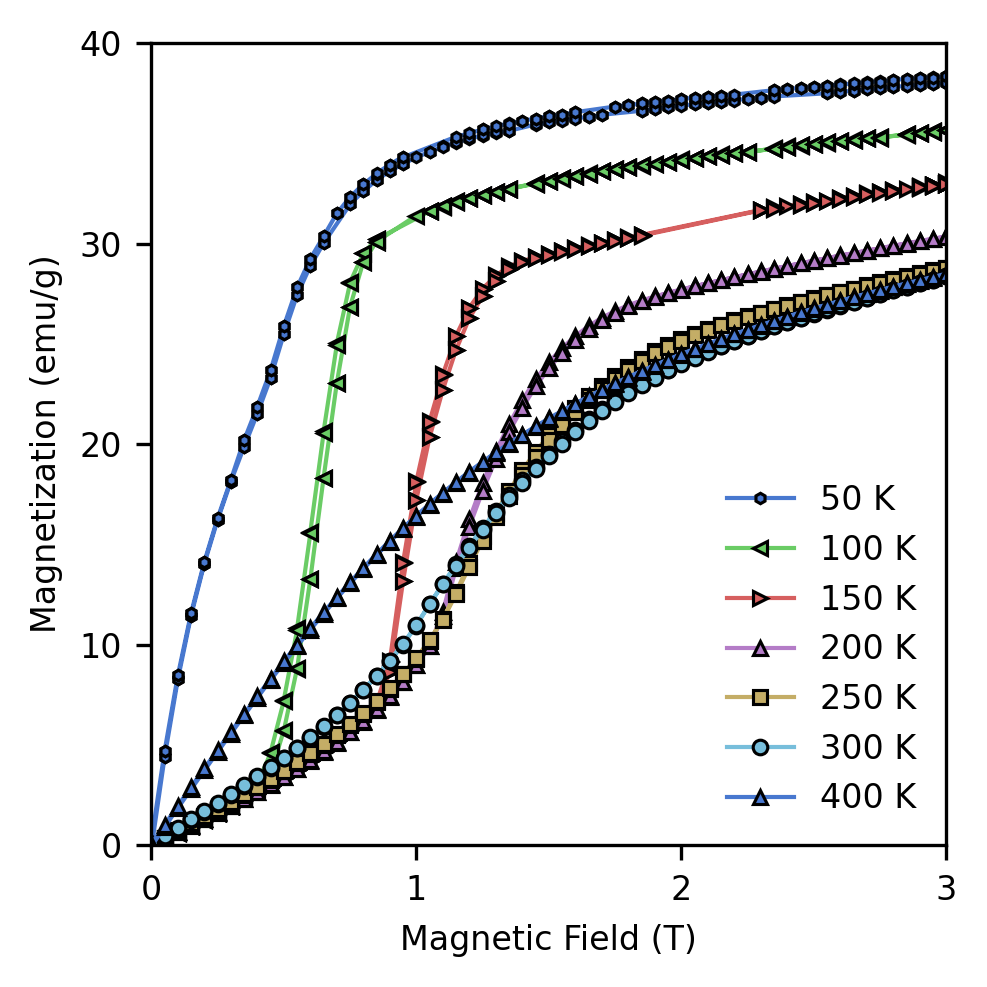}
    \caption{(right) Temperature-dependent magnetization of Fe$_{3}$Ga$_{4}$ samples annealed at 550 $^{\circ}$C with inset showing the derivative of magnetization with respect to temperature in the vicinity of the FM-ISDW (T$_{1}$) and ISDW-FM (T$_{2}$) magnetic transitions.  (left) Field-dependent magnetization of Fe$_{3}$Ga$_{4}$ samples annealed at 550 $^{\circ}$C from 0 - 3 T at various temperatures.} 
    \label{}
\end{figure*}

\begin{figure*}[]
    \centering
    \includegraphics[height = 3.2in, width = 3.2in]{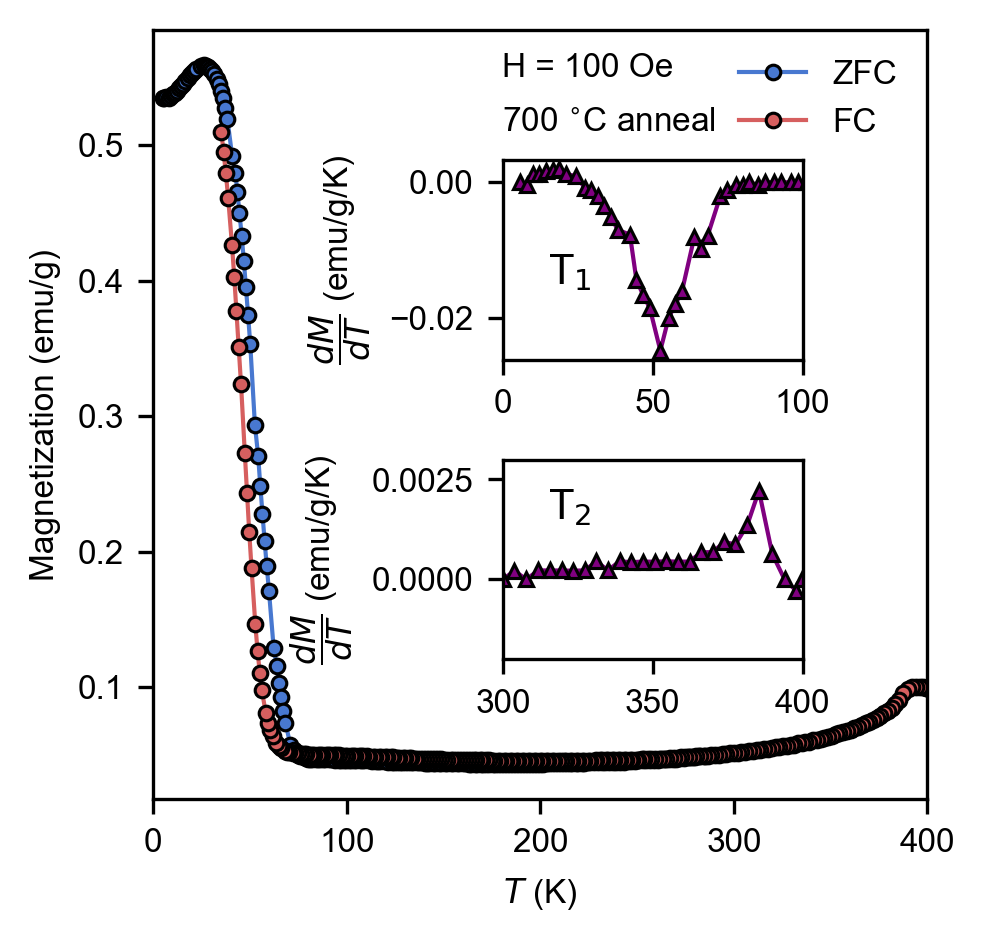}
    \includegraphics[height = 3.2in, width = 3.2in]{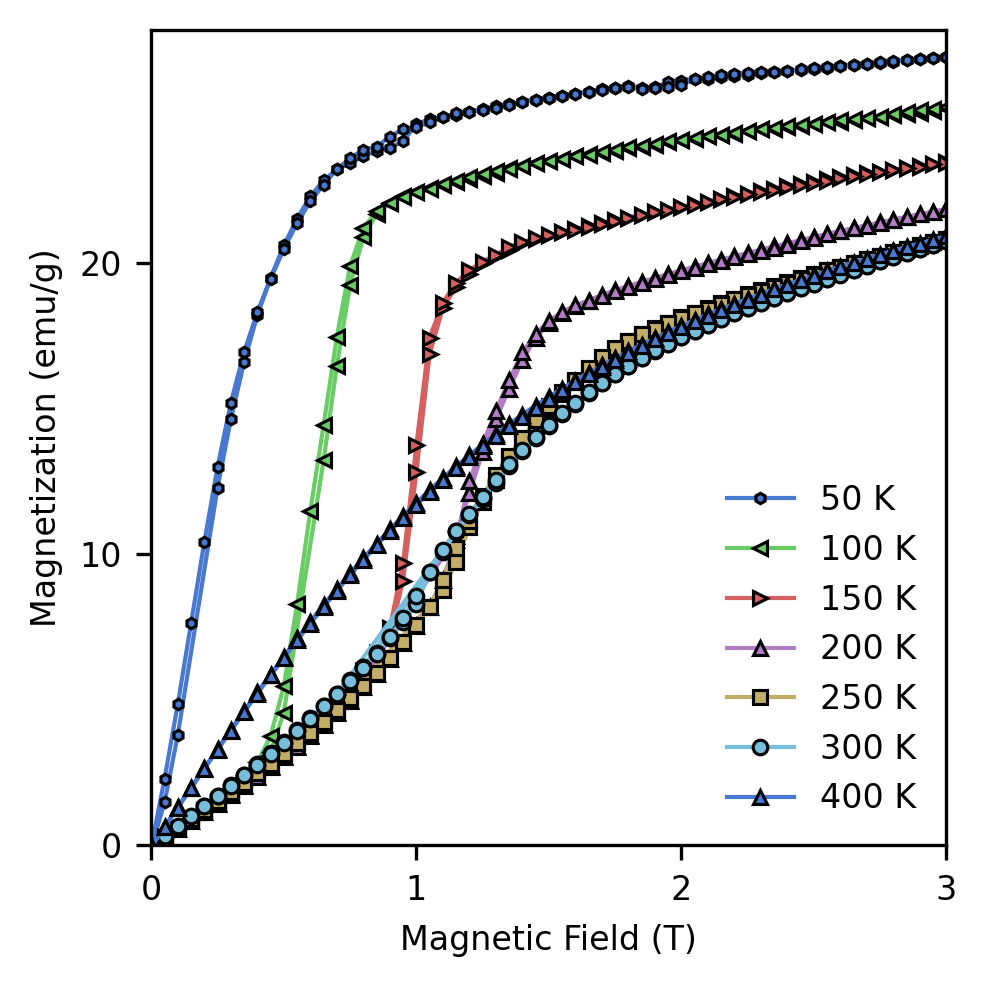}
    \caption{(right) Temperature-dependent magnetization of Fe$_{3}$Ga$_{4}$ samples annealed at 700 $^{\circ}$C with inset showing the derivative of magnetization with respect to temperature in the vicinity of the FM-ISDW (T$_{1}$) and ISDW-FM (T$_{2}$) magnetic transitions.  (left) Field-dependent magnetization of Fe$_{3}$Ga$_{4}$ samples annealed at 700 $^{\circ}$C from 0 - 3 T at various temperatures.} 
    \label{}
\end{figure*}

\begin{figure*}[]
    \centering
    \includegraphics[height = 3.2in, width = 3.2in]{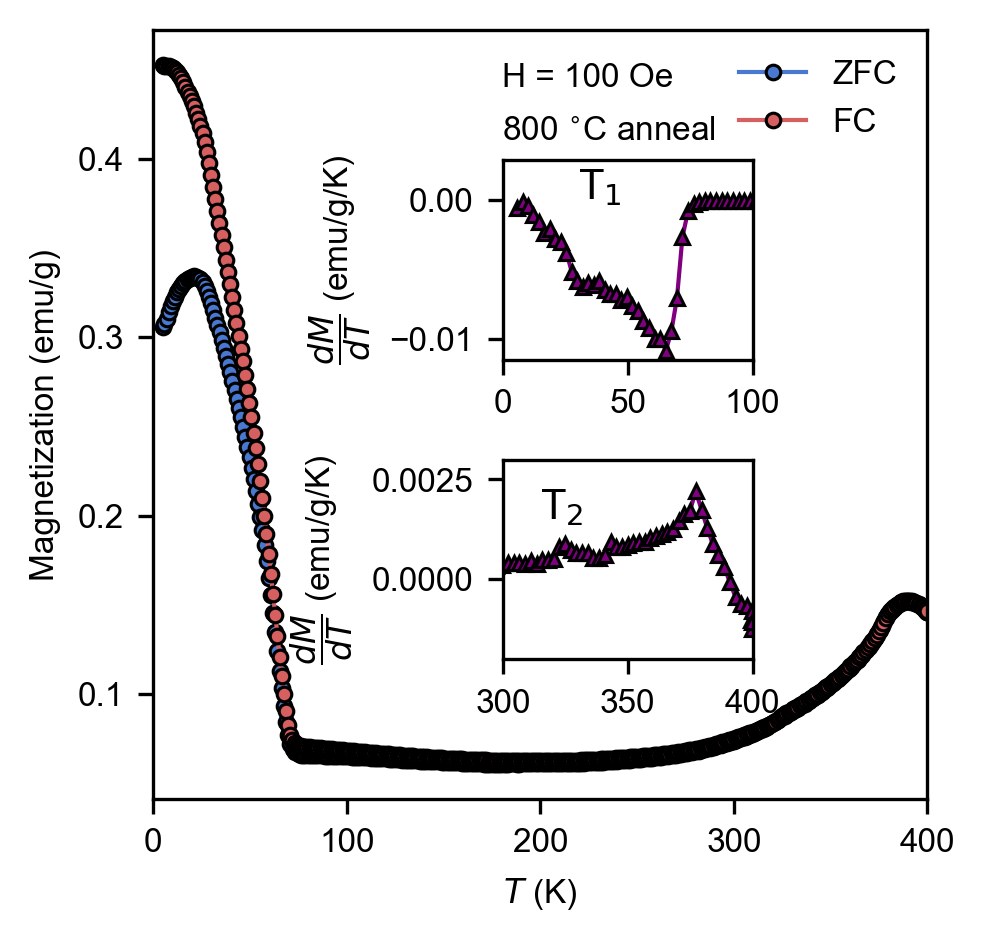}
    \includegraphics[height = 3.2in, width = 3.2in]{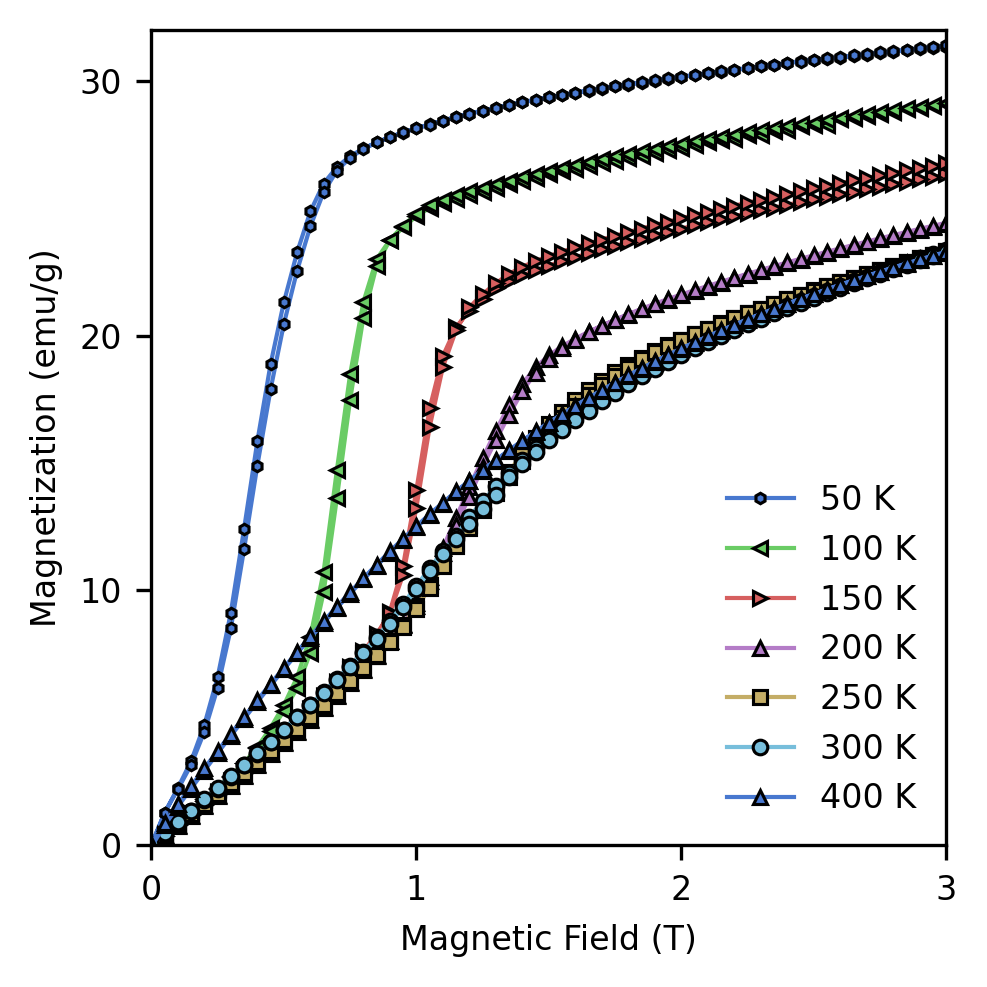}
    \caption{(right) Temperature-dependent magnetization of Fe$_{3}$Ga$_{4}$ samples annealed at 800 $^{\circ}$C with inset showing the derivative of magnetization with respect to temperature in the vicinity of the FM-ISDW (T$_{1}$) and ISDW-FM (T$_{2}$) magnetic transitions.  (left) Field-dependent magnetization of Fe$_{3}$Ga$_{4}$ samples annealed at 800 $^{\circ}$C from 0 - 3 T at various temperatures.} 
    \label{}
\end{figure*}

\begin{figure*}[]
    \centering
    \includegraphics[height = 3.2in, width = 3.2in]{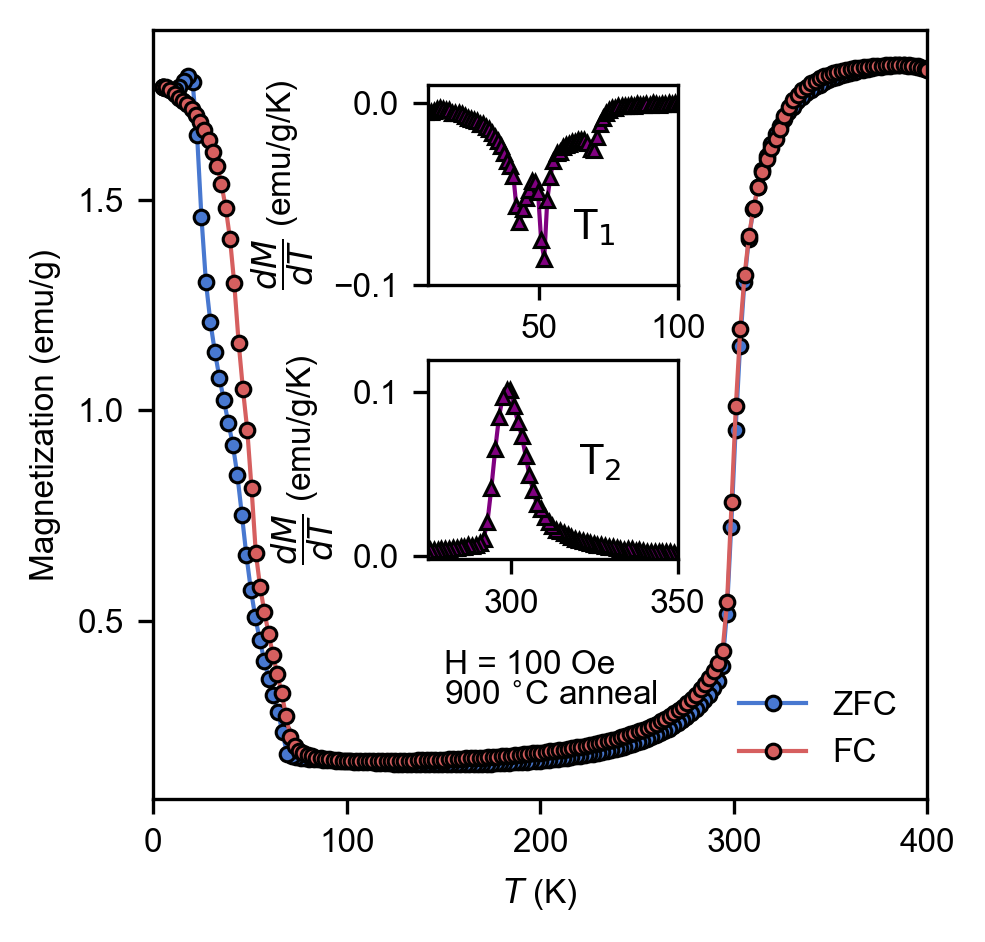}
    \includegraphics[height = 3.2in, width = 3.2in]{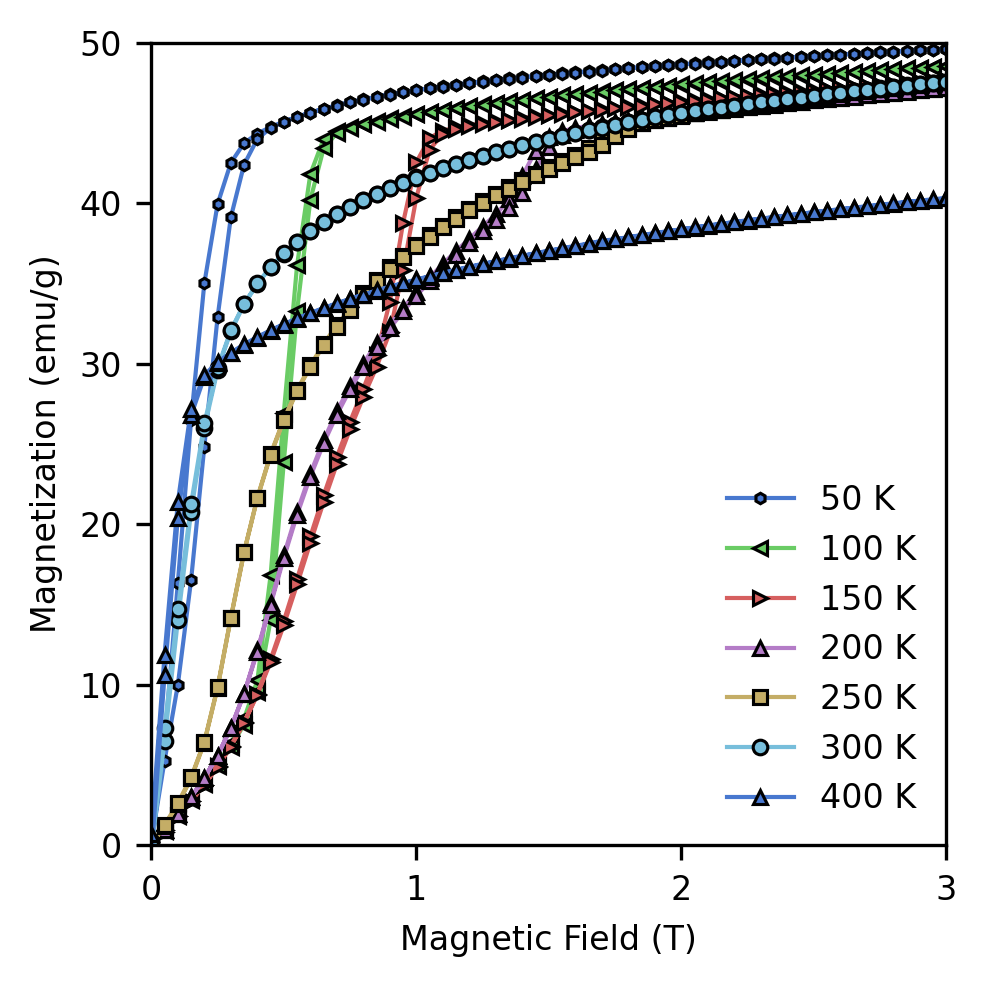}
    \caption{(right) Temperature-dependent magnetization of Fe$_{3}$Ga$_{4}$ samples annealed at 900 $^{\circ}$C with inset showing the derivative of magnetization with respect to temperature in the vicinity of the FM-ISDW (T$_{1}$) and ISDW-FM (T$_{2}$) magnetic transitions.  (left) Field-dependent magnetization of Fe$_{3}$Ga$_{4}$ samples annealed at 900 $^{\circ}$C from 0 - 3 T at various temperatures.} 
    \label{}
\end{figure*}

\begin{figure*}[]
    \centering
    \includegraphics[height = 3.2in, width = 3.2in]{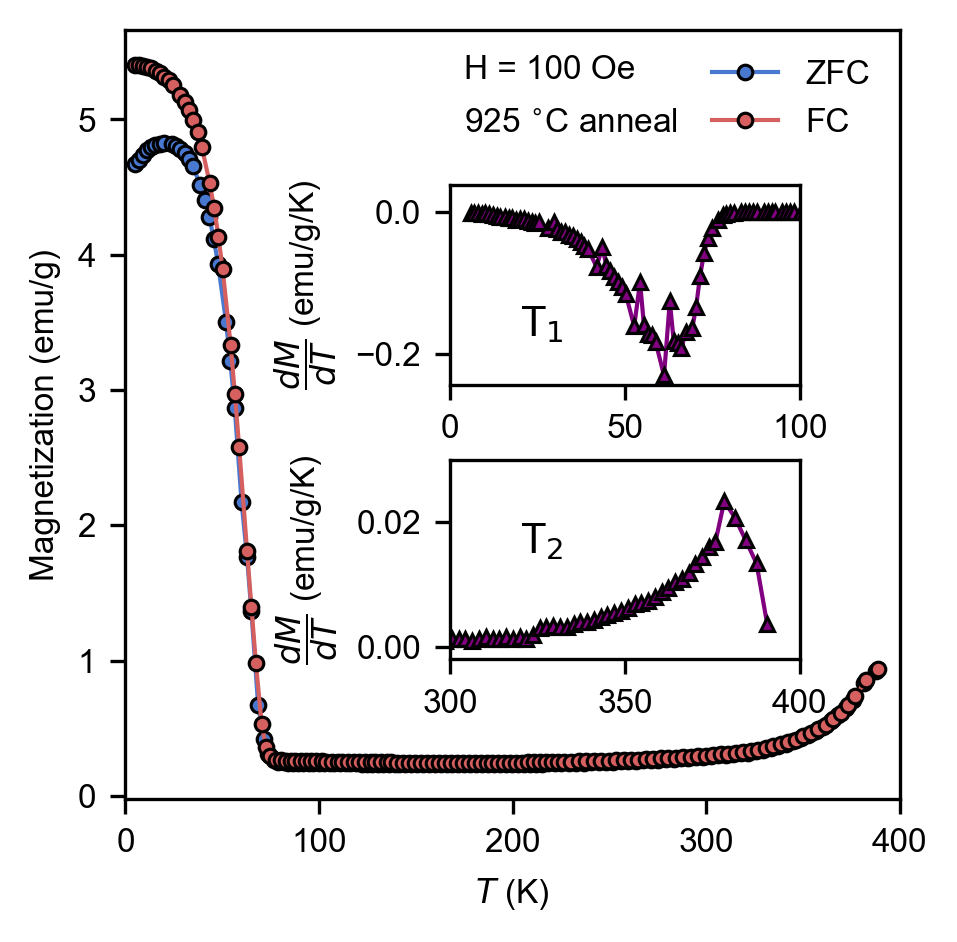}
    \includegraphics[height = 3.2in, width = 3.2in]{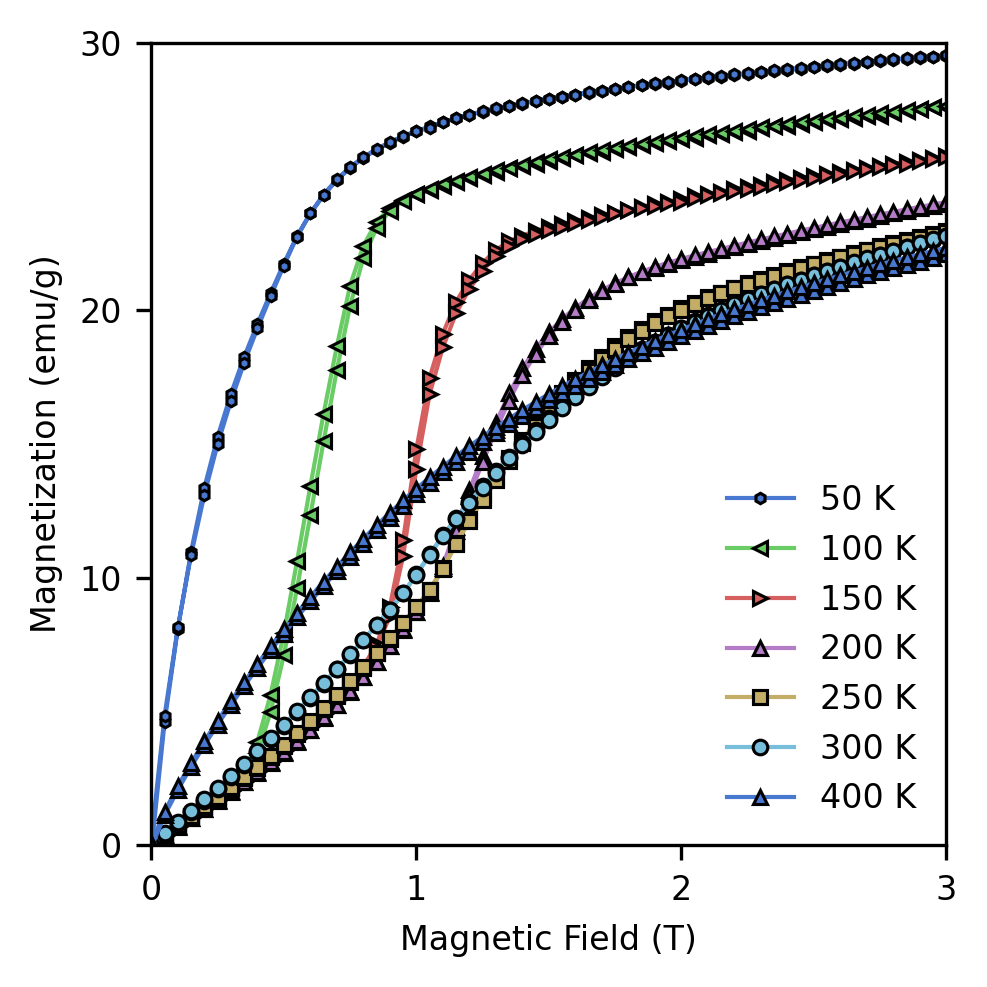}
    \caption{(right) Temperature-dependent magnetization of Fe$_{3}$Ga$_{4}$ samples annealed at 925 $^{\circ}$C with inset showing the derivative of magnetization with respect to temperature in the vicinity of the FM-ISDW (T$_{1}$) and ISDW-FM (T$_{2}$) magnetic transitions.  (left) Field-dependent magnetization of Fe$_{3}$Ga$_{4}$ samples annealed at 925 $^{\circ}$C from 0 - 3 T at various temperatures.} 
    \label{}
\end{figure*}

\begin{figure*}[]
    \centering
    \includegraphics[height = 3.2in, width = 3.2in]{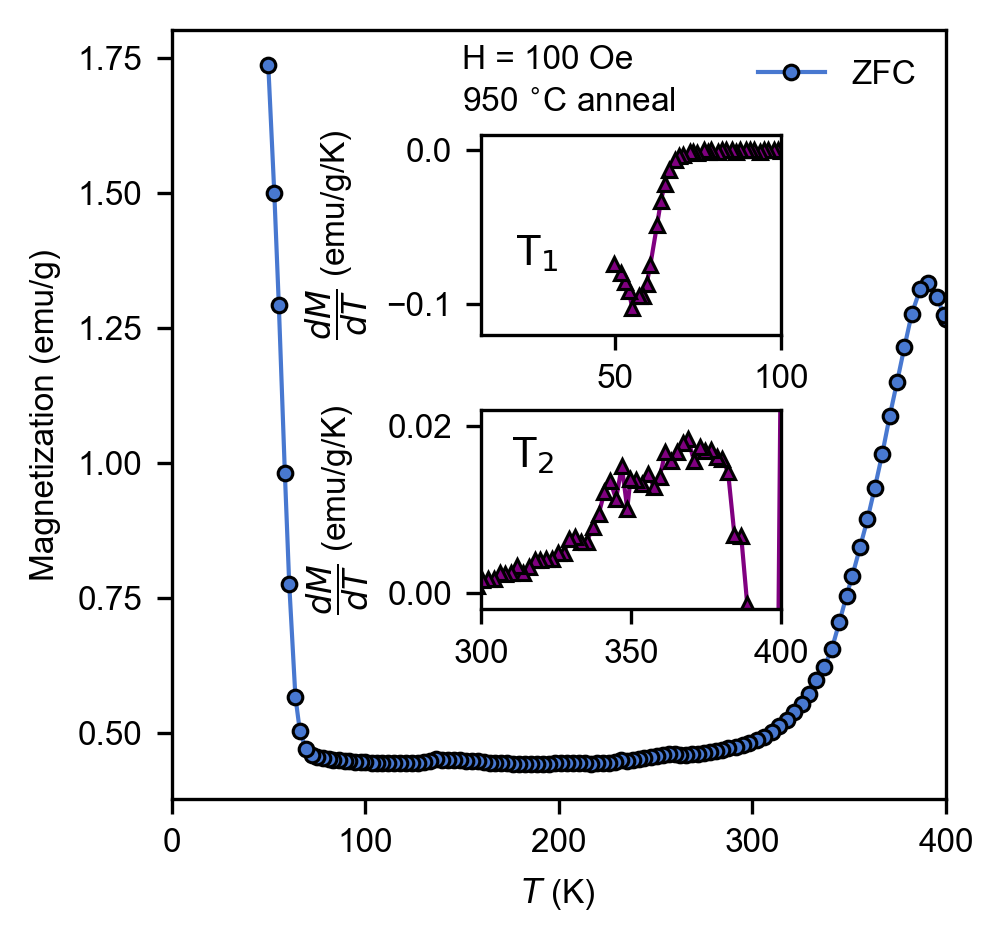}
    \includegraphics[height = 3.2in, width = 3.2in]{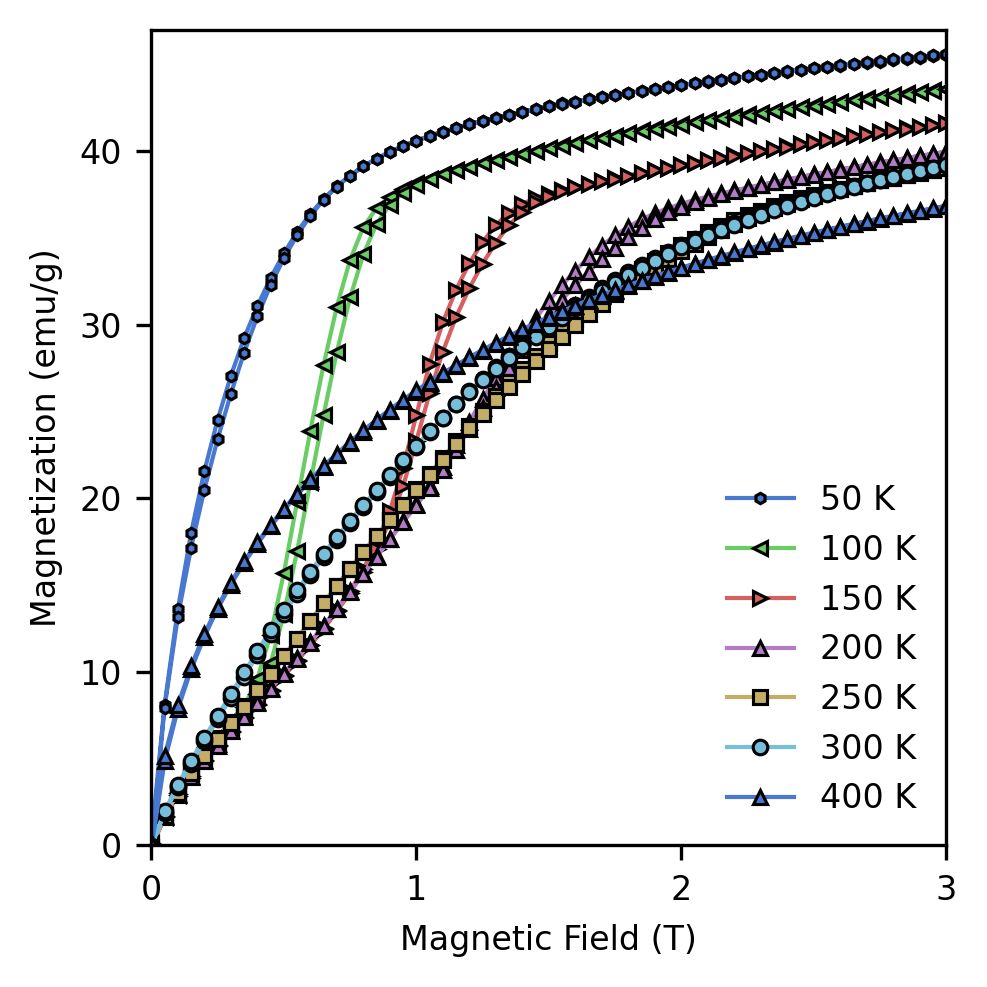}
    \caption{(right) Temperature-dependent magnetization of Fe$_{3}$Ga$_{4}$ samples annealed at 950 $^{\circ}$C with inset showing the derivative of magnetization with respect to temperature in the vicinity of the FM-ISDW (T$_{1}$) and ISDW-FM (T$_{2}$) magnetic transitions.  (left) Field-dependent magnetization of Fe$_{3}$Ga$_{4}$ samples annealed at 950 $^{\circ}$C from 0 - 3 T at various temperatures.} 
    \label{}
\end{figure*}

\begin{figure*}[]
    \centering
    \includegraphics[height = 3.2in, width = 3.2in]{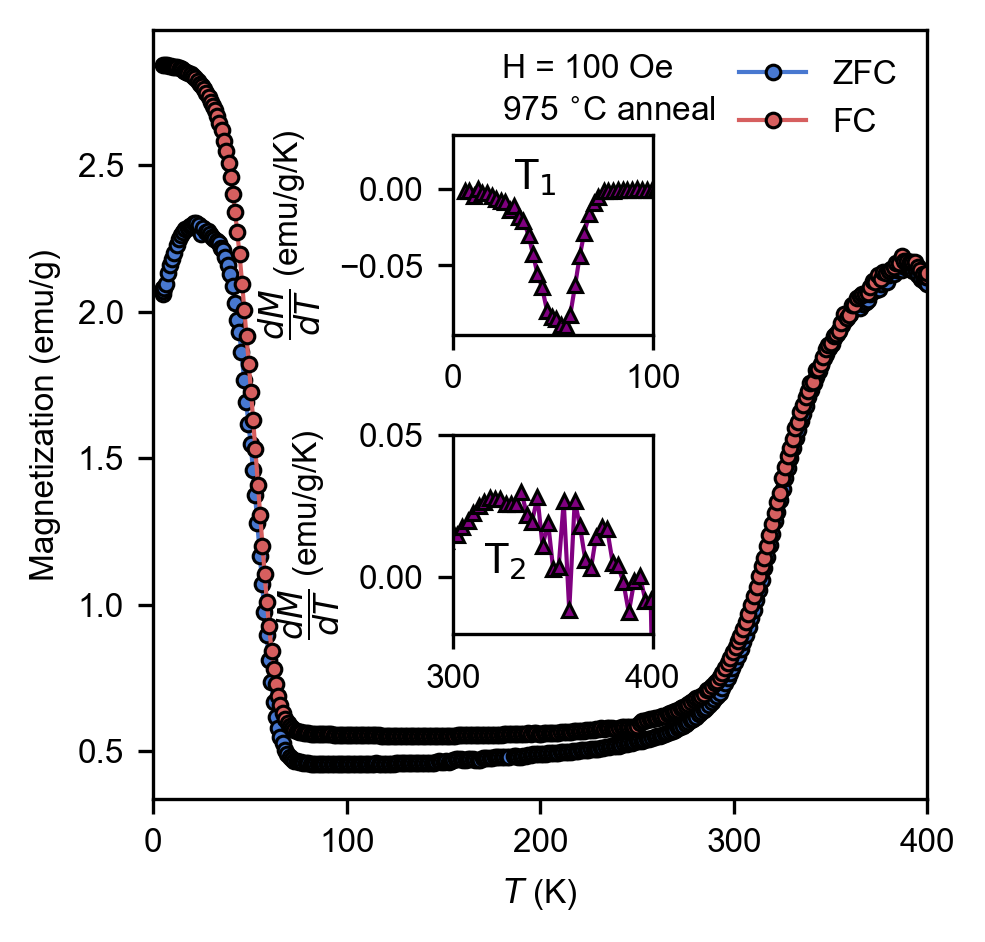}
    \includegraphics[height = 3.2in, width = 3.2in]{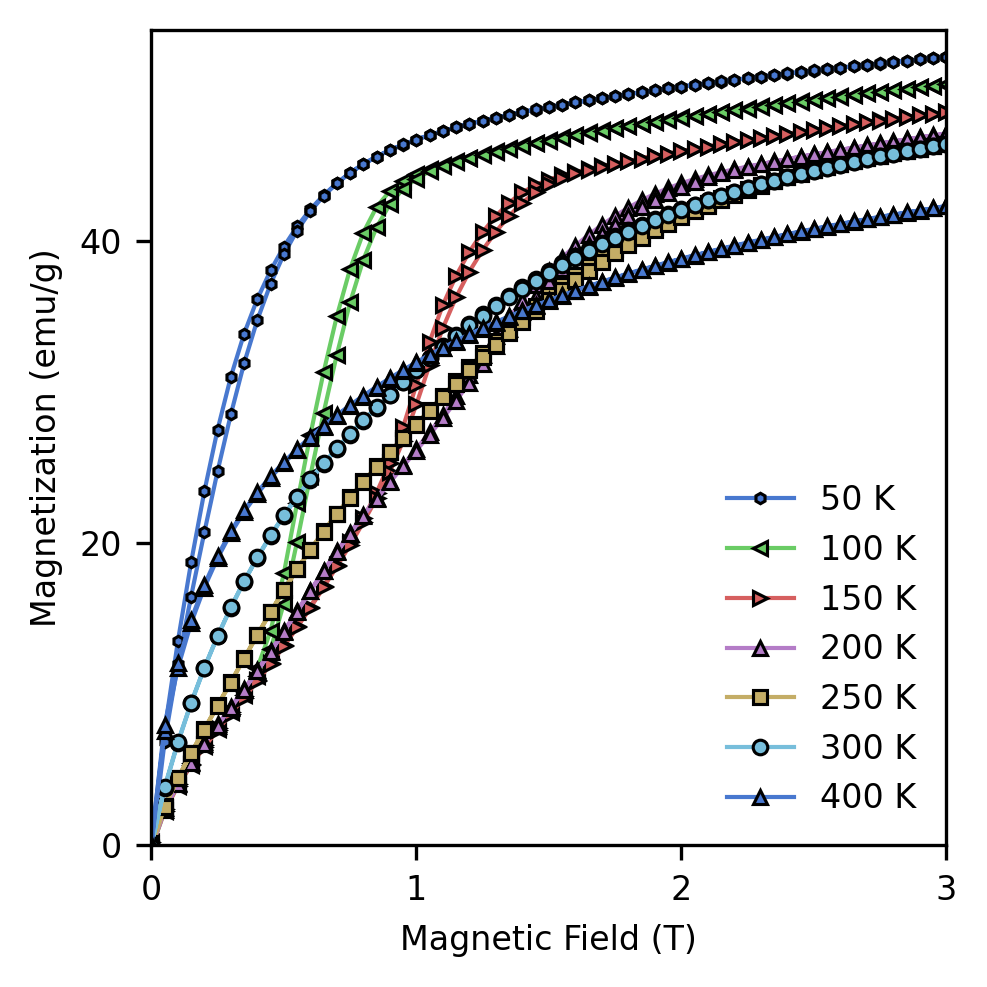}
    \caption{(right) Temperature-dependent magnetization of Fe$_{3}$Ga$_{4}$ samples annealed at 975 $^{\circ}$C with inset showing the derivative of magnetization with respect to temperature in the vicinity of the FM-ISDW (T$_{1}$) and ISDW-FM (T$_{2}$) magnetic transitions.  (left) Field-dependent magnetization of Fe$_{3}$Ga$_{4}$ samples annealed at 975 $^{\circ}$C from 0 - 3 T at various temperatures.} 
    \label{}
\end{figure*}

\begin{figure*}[]
    \centering
    \includegraphics[height = 3.2in, width = 3.2in]{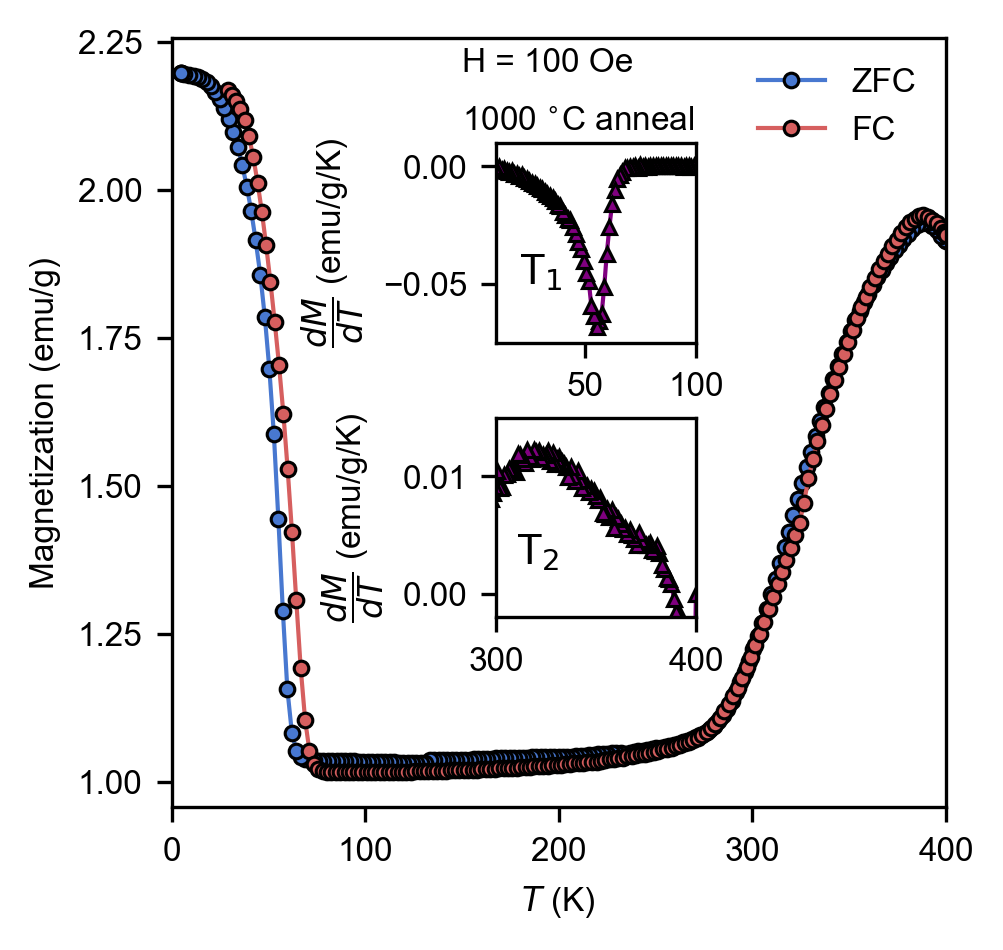}
    \includegraphics[height = 3.2in, width = 3.2in]{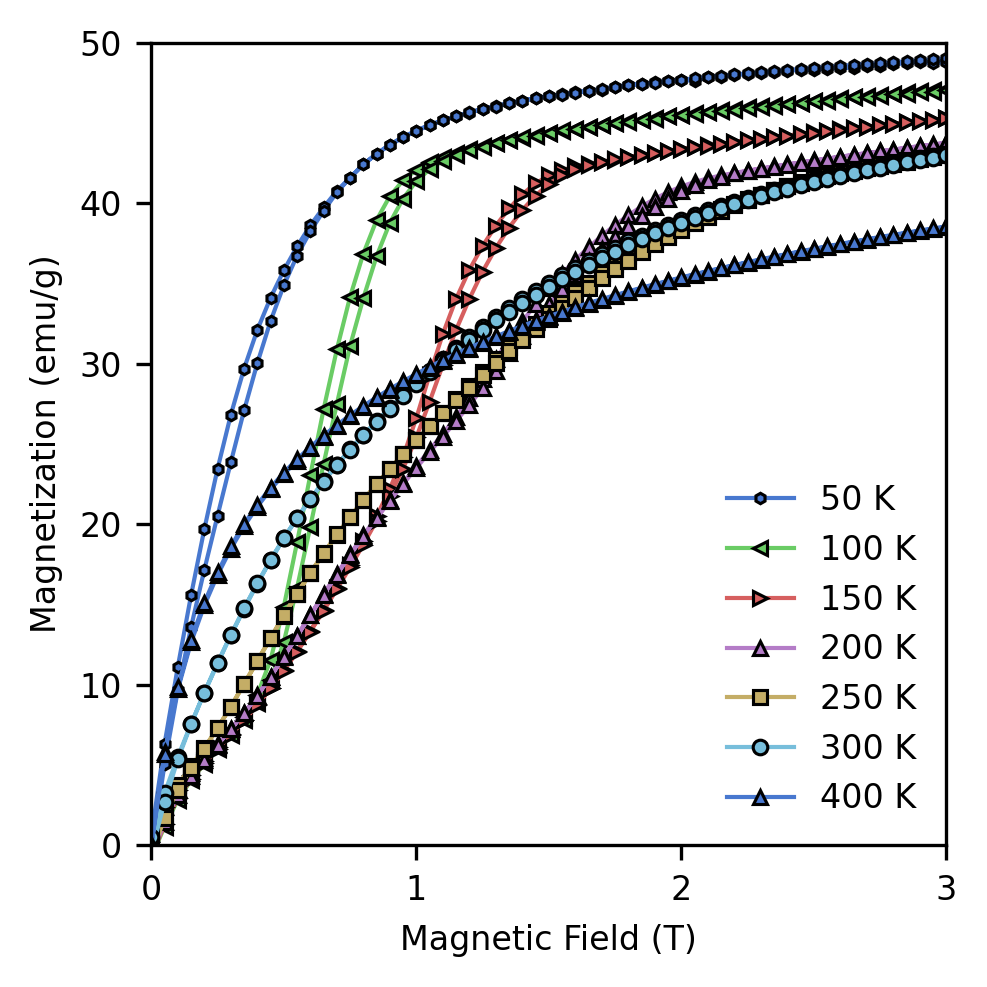}
    \caption{(right) Temperature-dependent magnetization of Fe$_{3}$Ga$_{4}$ samples annealed at 1000 $^{\circ}$C with inset showing the derivative of magnetization with respect to temperature in the vicinity of the FM-ISDW (T$_{1}$) and ISDW-FM (T$_{2}$) magnetic transitions.  (left) Field-dependent magnetization of Fe$_{3}$Ga$_{4}$ samples annealed at 1000 $^{\circ}$C from 0 - 3 T at various temperatures.} 
    \label{}
\end{figure*}

\end{document}